\documentclass[%
 aip,
amsmath,amssymb,
reprint,%
longbibliography,
]{revtex4-1}

\usepackage{graphicx}
\usepackage{dcolumn}
\usepackage{bm}

\usepackage[utf8]{inputenc}
\usepackage[T1]{fontenc}
\usepackage{mathptmx}
\usepackage{etoolbox}
\usepackage[colorlinks=true, allcolors=cyan]{hyperref}
\usepackage[nameinlink, noabbrev]{cleveref}

\usepackage{ragged2e}
\usepackage{caption}
\usepackage{subcaption}
\usepackage[normalem]{ulem}

\usepackage{xcolor}
\usepackage{balance}

\definecolor{myeditcolor}{rgb}{0, 0.5, 1.0}  

\DeclareCaptionJustification{forcejustify}{\justifying}

\crefname{figure}{Fig.}{Figs.}
\Crefname{figure}{Fig.}{Figs.}

\newif\ifHighlitedChanges
\def\ifHighlitedChanges{\iftrue}
\ifHighlitedChanges
   
  \def\STRIKE#1{{\color{red}\sout{#1}}}
\else

  \def\STRIKE#1{\relax}
\fi



\usepackage[resetlabels]{multibib}

\newcommand{\supprefname}{References} 
\newcites{S}{\supprefname}
\makeatletter
\def\@email#1#2{%
 \endgroup
 \patchcmd{\titleblock@produce}
  {\frontmatter@RRAPformat}
  {\frontmatter@RRAPformat{\produce@RRAP{*#1\href{mailto:#2}{#2}}}\frontmatter@RRAPformat}
  {}{}
}%
\makeatother
\begin{document}

\preprint{AIP/123-QED}
\title[]{Liquid-Liquid Phase Separation in a Minimal Explicit-Solvent Lattice Model Mimicking Protein Solutions}
\author{Siddhartha Roy}
 \affiliation{Department of Physics, Indian Institute of Science Education and Research (IISER) Tirupati, Tirupati, Andhra Pradesh 517619, India}
\author{Rakesh S. Singh${^*}$}
\email{rssingh@iisertirupati.ac.in}
\affiliation{Department of Chemistry, Indian Institute of Science Education and Research (IISER) Tirupati, Tirupati, Andhra Pradesh 517619, India}

\begin{abstract}
Biomolecular condensates play essential roles in cellular processes, and recent efforts have focused on understanding their assembly and rational design principles. In this study, we have employed an explicit-solvent minimal statistical mechanical model based on the lattice-gas Hamiltonian with quenched disorder --- which mimics crowders --- to investigate how protein-solvent and protein-crowder interactions influence condensate phase behavior and morphology. The computed phase diagrams reveal rich behavior, including upper critical solution temperature (UCST), closed-loop, and reentrant type transitions under varying protein-solvent interactions at both equilibrium and out-of-equilibrium conditions. We elucidated the origin of these phase behavior changes and examined the role of protein-crowder interactions in modulating condensed phase morphology and stability. We further extended this model to binary protein mixtures where we studied the phase behavior in the presence and absence of quenched disorder. Without disorder, the system exhibits diverse phase-separated morphologies --- partially wetted, fully wetted, segregative, and associative --- with phase boundaries delicately sensitive protein-solvent interactions. The introduction of quenched disorder (or crowder) leads to a broader spectrum of complex morphologies, dictated by the interplay among protein-protein, protein-solvent, and protein-crowder interaction parameters. In general, this work underscores that protein-solvent and protein-crowder interactions, together with protein-protein interactions, can act as key regulatory parameters for modulating condensate morphology. These insights may guide future computational and experimental studies of liquid-liquid phase separation in biomolecular systems aimed at designing stimuli-responsive condensates.
\end{abstract}

\maketitle

\section{INTRODUCTION}\label{sec I}
Biomolecular condensates (e.g. membraneless organelles), formed by liquid-liquid phase separation (LLPS) of macromolecules such as proteins and nucleic acids, are known to play significant roles in a wide range of important cellular functions~\cite{hyman_beyond_2012, hyman_liquid-liquid_2014, shin_liquid_2017, smokers_how_2024, das_tunable_2025, peeples_mechanistic_2021, riback_stress-triggered_2017, iserman_condensation_2020, palumbo_coevolution_2022, alberti_current_2025}. Recent research evidence suggests that aberrant forms of these condensates are associated with many human diseases, including cancer, neurodegeneration, and infectious diseases~\cite{wang_liquidliquid_2021, alberti_liquidliquid_2019, taylor_toxic_2002, cai_biomolecular_2021, song_unleashed_2026, mccormick_translation_2017, savastano_nucleocapsid_2020, cubuk_sars-cov-2_2021}. Exploration of the key mechanisms and factors that govern LLPS can pave the way for the advancement of therapeutic strategies for disease mitigation, as well as for bioengineering specific functional condensates. 

The functionalities of the condensates are governed by different environmental factors (or stimulus) such as the solvation medium, temperature ($T$), pressure ($P$), pH, salt and ion concentration, to name a few~\cite{moller_reentrant_2014, ruff_advances_2018, tanaka_exploring_2024, dignon_temperature-controlled_2019, adame-arana_liquid_2020, cinar_temperature_2019, riback_stress-triggered_2017, li_pressure_2021, cinar_pressureinduced_2018, joshi_hydrogen-bonded_2024, iserman_condensation_2020, krainer_reentrant_2021, alberti_current_2025}. There has been a growing interest in the characterization and comprehension of different condensate assemblies, along with the investigation of rational design principles that govern the constituents (like proteins), which exhibit a variety of stimulus-responsive phase behavior~\cite{quiroz_sequence_2015, farag_phase_2023, bremer_deciphering_2022, martin_valence_2020, dignon_sequence_2018, chew_thermodynamic_2023, rekhi_expanding_2024, das_relating_2015, benayad_simulation_2021, tesei_accurate_2021, zeng_design_2021, simon_programming_2017, chang_sequence_2017, andre_liquidliquid_2020, vweza_liquidliquid_2021, muthukumar_sticky_2026, ausserwoger_quantifying_2025, von_bulow_prediction_2025}. Also, studies reveal condensates with multi-phase morphology, where different dense phases forms layers around each other~\cite{feric_coexisting_2016, fare_higher-order_2021, regy_sequence_2020, kelley_amphiphilic_2021}. In terms of temperature as a stimulus, different types of $T$-dependent phase behavior of proteins are reported~\cite{dignon_temperature-controlled_2019, quiroz_sequence_2015, jiang_temperature-dependent_2025}. They can be broadly classified into three categories: (i) upper critical solution temperature (UCST) type, containing a demixed phase at low temperature but mixes at high temperature, (ii) lower critical solution temperature (LCST) type~\cite{quiroz_sequence_2015}, where proteins remain mixed at low temperature but condense above a certain temperature, and (iii) reentrant phase behavior, characterized by UCST at lower temperatures and LCST at higher temperatures, with a miscible region in between~\cite{dignon_temperature-controlled_2019}. The LCST-type  can lead to a closed-loop phase behavior with LCST at low temperature and UCST at high temperature~\cite{ruff_advances_2018} if the UCST is accessible in the explored $T$-range. Although UCST type systems are studied extensively leading to a relatively good understanding, the other two [LCST (including closed-loop) and reentrant] still remain relatively not-so-well explored and factors governing these types of anomalous phase behavior are a subject of active research (for example, see Ref.~\citenum{dhamankar_asymmetry_2024, chakravarti_accurate_2025}). 

Furthermore, water constitutes approximately $70\%$ of the cytosol, and its role as a solvent in modulating intracellular LLPS is an additional avenue of current investigation~\cite{joshi_hydrogen-bonded_2024, mukherjee_therodynamic_2023, mukherjee_entropy_2024, thirumalai_role_2012, ribeiro_synergic_2019, ahlers_2021, pezzotti_2023}. Nevertheless, the precise mechanisms by which the solvent modulates LLPS --- particularly through the lens of conformation-dependent protein hydration and inter-protein interactions --- remain poorly understood. In addition to solvent, LLPS and the corresponding condensate morphologies can also be affected by the fact that the cytosol is a complex medium containing several categories of macromolecules or crowders (such as proteins, nucleic acids, polysaccharides, etc). The resultant dense~\cite{andre_liquidliquid_2020, vweza_liquidliquid_2021} and highly heterogeneous environment underscores the necessity for a comprehensive investigation of the effects of inhomogeneity (crowder) in modulating LLPS under equilibrium and out-of-equilibrium (which arises from active processes and is intrinsic to the intracellular medium) conditions. Therefore, investigating these aspects requires a robust understanding of the interplay between solvent, intracellular inhomogeneity and non-thermal noise-driven effects on the alterations in a protein's (or heteropolymer's) conformational state to the resulting changes in inter-protein interactions and in turn their phase behavior~\cite{garaizar_expansion_2020, rovigatti_designing_2022, rovigatti_entropy-driven_2023}. Establishing this linkage remains highly challenging because it requires the simultaneous exploration of molecular- and mesoscale-resolution phenomena, which is difficult to achieve in both experiments and computational studies. Computational approaches often rely on implicit-solvent coarse-grained protein models, while experimental probes are limited by spatial resolution in directly resolving local interactions. Such multiscale linkages, however, can provide an extensive framework of design principles governing homotypic and heterotypic interactions, which are likely essential for the formation and functionality of biomolecular condensates~\cite{chakravarti_accurate_2025, boeynaems_protein_2018, kim_thermo-responsive_2017, dai_engineering_2023, tamaki_ph-switchable_2020, tangade_multiphasic_2025}. 
 
With the aim of understanding the role of conformation dependent protein-solvent and protein-inhomogeneity (crowder) interactions in biomolecular condensate behavior, in this study we have constructed a minimal lattice-gas (LG) model accounting explicitly for solvents and inhomogeneity, the latter being implemented as quenched disorder. This model improves upon our previous implicit solvent model for the LLPS of proteins~\cite{roy_anomalous_2025} by considering more realistic the factors associated with intracellular environment. Since proteins are capable of adapting multiple conformational states (in contrast to simple molecules), the interactions between them depends on their internal conformational state which can be altered through solvent properties or other external factors (including thermodynamic condition). In turn the protein-solvent and protein-crowder interaction can also be greatly sensitive to the protein's internal conformation state. Therefore, the sequence composition, along with folding-unfolding thermodynamics --- which governs conformational population --- is recognized as a pivotal factor contributing to protein aggregation and phase separation~\cite{dignon_temperature-controlled_2019, lichtinger_targeted_2021, rekhi_expanding_2024, sahli_role_2019, garaizar_expansion_2020, nikfarjam_effects_2020, zeng_connecting_2020, dannenhoffer-lafage_data-driven_2021, doi:10.1021/acs.jpclett.9b01731, hazra_biophysics_2021, regy_sequence_2020, kaur_sequence-encoded_2021, dignon_relation_2018}. 

Our model mimics the proteins with particles having multiple internal states representing different conformational states. For the sake of simplicity we have considered two internal states: a non-degenerate ground state and a degenerate excited state representing two classes of protein conformational ensemble significantly differing in their extent of folded (or unfolded) domains. The solvents are mimicked by particles representing a volume element corresponding to a localized fluid volume of scale similar to the excluded volume of a protein unit. The static disorder/impurity is mimicked by cluster of pinned particles, which only interact with other species like proteins and solvent. Our results shed light on the anomalous reentrant behavior along with capturing exclusive UCST and closed loop (including LCST) behavior separately under varied protein-solvent interactions at both equilibrium and non-equilibrium conditions. We further extended our study to a binary protein system and also probed the effects of crowders on the condensate morphology and stability. Our results may help in providing key insights into the design of protein sequences and solvent (including other environmental factors such as crowder) for targeted biomolecular condensate morphology and stability.  

The layout of the subsequent sections of this paper is as follows. Sec.~\ref{secII} introduces our lattice model with explicit solvent and disorder along with discussing the details about the Monte-Carlo methods used to simulate this model system. Secs.~\ref{sec_IIIA} and~\ref{sec_IIIB} report the protein-solvent interaction induced alteration of the phase diagram (ranging from reentrant, UCST and close-loop phase diagrams), followed by origin of the same for the single component protein system with conformational interconversion. In Sec.~\ref{sec_IIID}, we discuss the modulation of LLPS by protein-solvent interactions in a binary protein system where conformational interconversion is not allowed. Sections~\ref{sec_IIIC} and~\ref{sec_IIIE} describe the effects of quenched disorder on phase separation in both single component and binary protein systems, respectively. The conclusions from this work are summarized in Sec.~\ref{sec_IV}.
\begin{figure*}[!t]
    \centering
    \begin{minipage}{\textwidth}
    \includegraphics[width=0.96\textwidth]{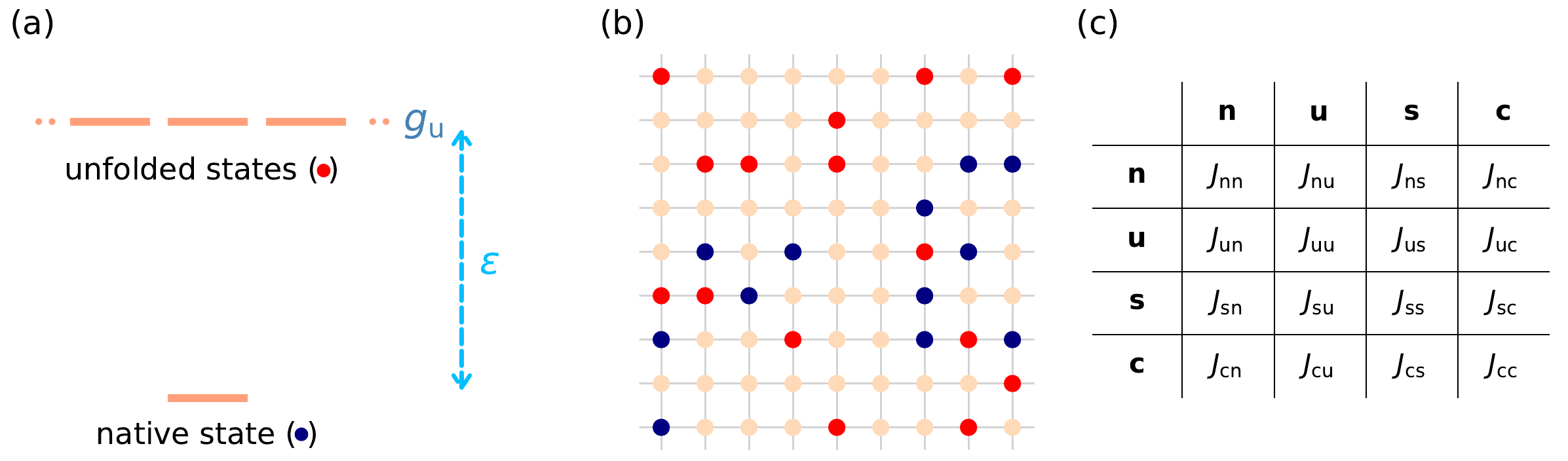}\label{fig1}
     \phantomsubcaption\label{fig_1a}
     \phantomsubcaption\label{fig_1b}
     \phantomsubcaption\label{fig_1c}
    \end{minipage}
   \caption{(a) Proteins are represented with particles having two internal states, native (n, denoted by a blue filled circle) and unfolded (u, denoted by a red filled circle). The degeneracies of the same are $1$ and $g_{\rm u}$, respectively. In the context of IDPs, as they lack a well-defined stable native structure, these two states faithfully represent two classes of protein conformations with significantly varying degrees of folded domains. $\epsilon$ is the internal interaction energy difference between the two conformational states that accounts for the loss of internal weak interactions on conformational transition. (b) Schematic representation of the lattice system with randomly dispersed native and unfolded proteins along with solvents (filled peach colored particles). (c) Interaction matrix representing the different kinds of interaction parameters involving proteins, solvents and static impurities in our minimal model.}
\end{figure*}
\section{MODEL AND METHOD DETAILS}\label{secII}
\subsection{Lattice-gas model mimicking proteins in solvent }\label{secIIA}
 In our lattice model, we represent proteins by particles with multiple internal states, a non-degenerate ground state and degenerate higher energy states corresponding to the native and unfolded protein configurations respectively (see Fig.~\ref{fig_1a}) --- akin to our previous implicit solvent model proposed in Ref.~\citenum{roy_anomalous_2025}. Although, for the sake of simplicity we designate these states as ``native'' and ``unfolded'' (denoted by `n' and `u', respectively), in the context of IDPs (characterized by a heterogeneous ensembles of conformations with a lack of a well-defined native structure under physiological conditions), they more faithfully represent two classes of protein conformations with significantly varying degrees of folded domains. The solvents are represented by particles with a single internal state. Since a protein is much larger in size (hydrodynamic radius) than a typical solvent molecule (say water), our framework represents a ``coarse-graining'' of the solvent where each lattice site designated as ``explicit solvent'' is defined not as a discrete molecular entity, but as a mesoscopic volume element --- a solvent packet --- representing a localized fluid volume equivalent to the excluded volume of a protein unit. Thus, our model does not resolve solvent density fluctuations at scales smaller than this length scale which can be much larger than the characteristic density correlation length of the solvent. The static inhomogeneity (or quenched disorder) in the system is introduced by clusters of pinned particles (fixed on lattice sites) that interact with proteins --- differently for different conformational states --- and with the solvent.
 
 We use a two-dimensional (2D) square lattice consisting of $N$ sites, with each site characterized by the occupation number $n$, which can assume a value of either $1$ or $0$, indicating the presence or absence of a particle (protein or solvent) at site $i$, respectively (see Fig.~\ref{fig_1b}). Additionally, the internal states of the particles are represented by a variable $\sigma$, which can take on $p$ discrete values, specifically $1$, $2$, $\ldots$, $p$; where $p$ denotes the total number of internal conformational states of the protein. In this work, we have chosen $p =2$ for proteins and for other species (i.e. solvent and disorder particles), $p= 1$. The Hamiltonian of the system is given by: 
\begin{equation}\label{eq2.1}
\begin{split}
    H = & - \sum_{ \langle ij \rangle} J_{\rm pp}( \sigma_i, \sigma_j ) n_i n_j - J_{\rm ss} \sum_{ \langle ij \rangle} l_i l_j - \sum_{ \langle ij \rangle} J_{\rm ps}( \sigma_i) n_i l_j \\
        &  - \sum_{ \langle ij \rangle} J_{\rm cp}( \sigma_i) n_i m_j - J_{\rm cs} \sum_{ \langle ij \rangle} m_i l_j  + \sum_{i} \epsilon( \sigma_i) n_i. 
\end{split}
\end{equation}
Here, $n_i$, $m_i$ and $l_i$ denote the occupancy index of the $i^{\rm th}$ site for protein, impurity particle and solvent, respectively. Also, $J_{\rm pp}( \sigma_i, \sigma_j )$, $J_{\rm ps}( \sigma_i)$ and $J_{\rm cp}( \sigma_i)$ (with p = u and n) represent the inter-protein, protein-solvent and protein-impurity cluster interactions, respectively, which dependent on the protein's internal state. $J_{\rm ss}$ and $J_{\rm cs}$ correspond to the inter-solvent and cluster-solvent interactions, respectively (see the interaction matrix shown in Fig.~\ref{fig_1c}) and $\langle ... \rangle$ denotes the nearest neighbor interaction between different or same species. We have set the internal energy of the native proteins to $\varepsilon (1) = 0$ and that of the unfolded proteins to $\varepsilon (2) = 2.0J_{\rm uu}$~\cite{roy_anomalous_2025}. It is worth noting here that several minimal lattice models have been previously employed to study phase behavior under both equilibrium and non-equilibrium conditions, albeit in contexts different from those addressed in this study~\cite{caupin_minimal_2021, longo_structure_2022, longo_phase_2022, longo_formation_2023, cho_tuning_2023, cho_nonequilibrium_2023}.

In the present study, since our goal is to explore the role of protein-solvent and protein-impurity (or, crowder) interactions on LLPS --- the homotypic protein-protein interactions $J_{\rm pp}$ ($J_{\rm uu}$, $J_{\rm nn}$ for p = u and n), the solvent-solvent interactions ($J_{\rm ss}$) and the cluster-solvent interactions ($J_{\rm cs}$) are kept fixed. Specifically, the parameters are judiciously set as follows: $J_{\rm uu} = 1.0$, $J_{\rm nn} = 0.5$, $J_{\rm ss} = 0.25$, $J_{\rm cs} = 0.15$. The condition $J_{\rm uu} > J_{\rm nn}$ accounts for the loss of available interaction sites due to internal burial in the folded state. We have set the unfolded state degeneracy $g_{\text{u}}$ to $20$. This parameter is chosen so that, for the given inter-protein and protein-solvent interaction parameters, a significant conformational transition is observed within the temperature range probed here. We explore the phase behavior of our system by varying the heterotypic interaction, $J_{\rm un}$, $J_{\rm ps}$ and $J_{\rm cp}$ (with p = u and n). The choice of $J_{\rm ps}$ (with p = u and n) is made based on the calculation of the Flory-Huggins $\chi$ parameter \cite{flory_thermodynamics_1942, huggins_thermodynamic_1942} at different $T$ values. The values are chosen to tune the miscibility of the proteins, thereby facilitating the exploration of various demixed and mixed phases under a range of thermodynamic conditions. In the subsequent sections, we have systematically investigated the effects of protein-solvent interaction on the phase behavior and condensate morphology for different scenarios representing in and out-of-equilibrium situations as well as in the presence and absence of static impurities. 
\begin{figure*}[htbp!]
    \centering
    \begin{minipage}{\textwidth}
    \includegraphics[width=\textwidth]{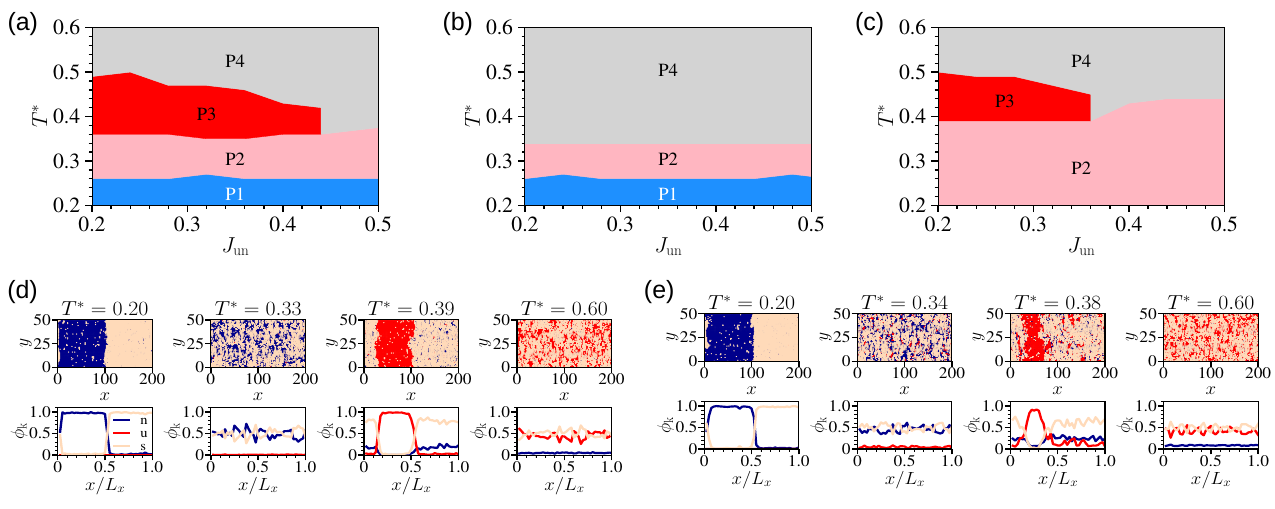}
     \phantomsubcaption\label{fig_2a}
     \phantomsubcaption\label{fig_2b}
     \phantomsubcaption\label{fig_2c}
     \phantomsubcaption\label{fig_2d}
     \phantomsubcaption\label{fig_2e}
    \end{minipage}
   \caption{Phase diagram in the $T^{*} - J_{ \rm un}$ plane for different protein-solvent interactions ($J_{\rm ps}$ with p = u and n). Here, the protein fraction $\phi_{\text{p}}$ is set to $0.5$ and protein-protein ($J_{\rm nn}$ and $J_{\rm uu}$) and solvent-solvent ($J_{\rm ss}$) interaction parameters are set at the values reported in Sec.~\ref{secIIA}. (a) The reentrant phase behavior is observed for $J_{ \rm us} = 0.20$ and $J_{\rm ns} = 0.15$. Here, a predominantly native phase-separated state P1 undergoes a mixing transition to P2 with increasing $T$ and then further transitions to an unfolded-dominated demixed state P3. On raising the temperature to sufficiently high values, the systems mixes again into a homogeneous state, mostly comprising of unfolded state proteins. The representative snapshots along with the composition profiles for state points belonging to P1 - P4 phases are shown in (d) and (e) for $J_{\rm un}$ = 0.20 and $J_{ \rm un}$ = 0.44, respectively. The blue, red and peach colors in the snapshots represents native protein, unfolded protein and solvent, respectively. The P1-P2 boundary marks the UCST-I line, while the P2-P3 and P3-P4 boundaries mark the LCST and UCST-II lines, respectively. (b) Exclusive UCST behavior is observed for $J_{\rm us} = 0.35$ and $J_{\rm ns} = 0.15$, where the mixed native-dominated P2 phase gradually transitions to unfolded-dominated P4 phase without the intermediate demixed P3 phase. In this phase diagram we report a single UCST line separating the P1 and P2 phases. (c) A closed-loop phase diagram for $J_{\rm us} = 0.20$ and $J_{\rm ns}$ = 0.25. Here, the LCST line separates P2 (mixed) and P3 (demixed) phases and UCST line separates P3 (de-mixed) and P4 (mixed) phases. The native-dominated phase separated phase P1 disappears from the phase diagram in the reported temperature range.}
   \label{fig2}
\end{figure*} 
\subsection{Computational methods}\label{secIIB}
We employed Monte-Carlo simulations in the canonical ensemble to sample the configuration space, characterized by conformation interconversion dynamics of proteins and particle (both protein and solvent) translation across the lattice (see Ref.~\citenum{roy_anomalous_2025} for details). To study the phase behavior, we consider two types of order parameters: the total fraction of proteins ($\phi_{\rm p}$, a conserved variable) and the fraction of proteins in the unfolded state ($\phi_{\rm u}$; note $\phi_{\rm n} = 1- \phi_{\rm u}$), a non-conserved variable. In contrary to our previous study~\cite{roy_anomalous_2025}, all the lattice sites are occupied here either by proteins, solvents and impurity clusters. The translation of proteins and solvents on the lattice is governed by Kawasaki exchange dynamics~\cite{PhysRev.145.224}, and the proteins' conformational space is sampled through excitation de-excitation dynamics with Metropolis algorithm. In Kawasaki exchange dynamics, the positions of a pair of proteins (with different internal states) or a protein-solvent pair are exchanged. On the other hand, the particles in the static impurity (or, disorder) clusters are fixed throughout the duration of the simulation, with the peripheral particles interacting with proteins and solvents. In both the cases (conformational interconversion of proteins or translation of proteins and solvents), the conversion or exchange dynamics depends on the energy difference ($\Delta E$) involved in the process, with the acceptance probability of a proposed move is described by the Metropolis criterion~\cite{frenkel_understanding_2023}, represented as $p_{\rm acc} = \text{min}[1, e^{-\beta \Delta E}]$; $\beta = 1/k_{\rm B}T$. We employed square and rectangular simulation box with periodic boundary conditions along both $x$ and $y$ axes. The lattice sizes ($L_x \times L_y$) used for square and rectangular boxes are $100 \times 100$ and $200 \times 50$, respectively. In the  forthcoming sections, temperature is reported in units of $J_{\rm uu}/k_{\rm B}$ as $T^* = k_{\rm B}T/J_{\rm uu}$. 
 \begin{figure*}
    \centering
      \begin{minipage}{0.8\textwidth}
    \includegraphics[width=\linewidth]{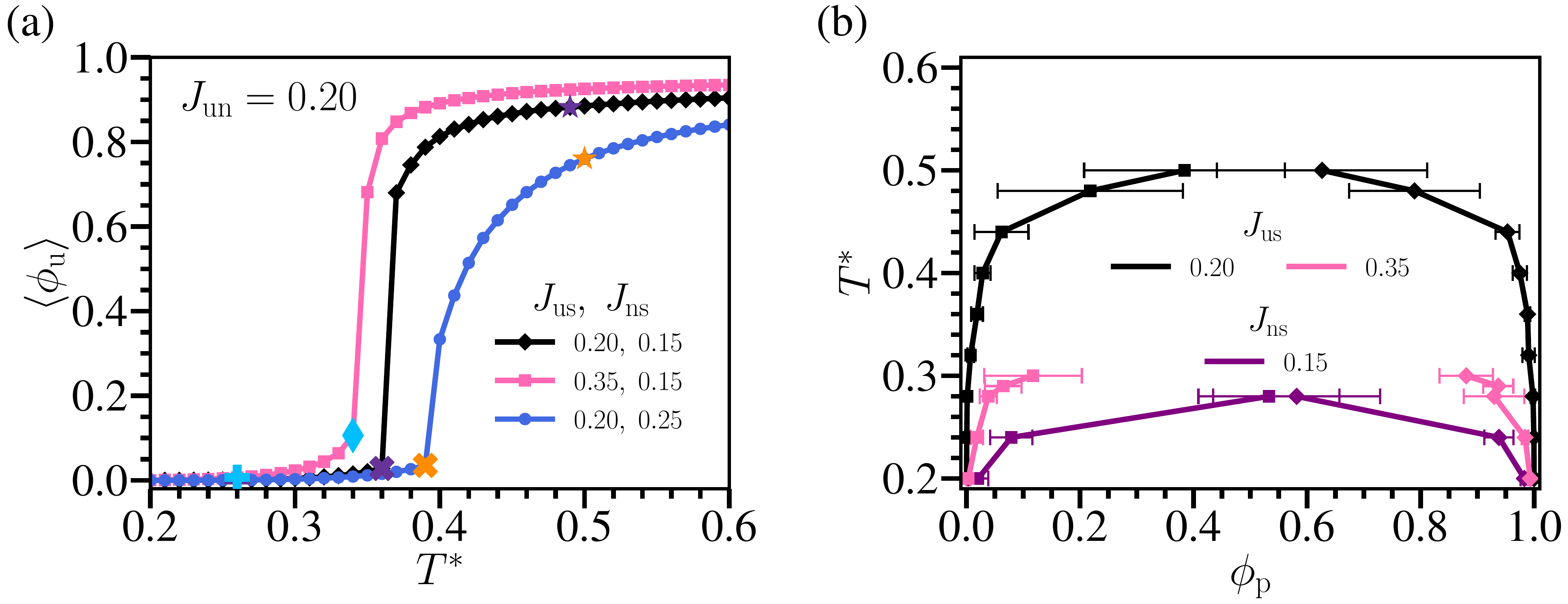}
     \phantomsubcaption\label{fig_3a}
     \phantomsubcaption\label{fig_3b}
         \end{minipage}
    \caption{(a) Temperature dependence of average fraction of unfolded state proteins ($\langle \phi_{\rm u} \rangle$) for protein-solvent interactions ($J_{\rm ps}$ with p = u and n) corresponding to three different scenarios presented in Figs.~\ref{fig_2a}-\ref{fig_2c} --- reentrant (black line) , UCST (pink line) and closed-loop (blue line) type phase behavior --- for a fixed heterotypic interaction ($J_{\rm un} = 0.20$). (b) Phase diagram in the $T^* - \phi_{\rm p}$ plane where all the proteins are either in the unfolded state or the native state for different values of protein-solvent interactions. We observe that the phase diagram is highly sensitive to the choice of these interactions; specifically, increasing the interaction strength shifts the critical solution temperature to a lower value.}
    \label{fig3}
\end{figure*}
\section{Results and Discussion}\label{sec III}
\subsection{Effects of protein-solvent interaction on LLPS}\label{sec_IIIA}
In Figs.~\ref{fig_2a}-\ref{fig_2c}, we report the $T - J_{\rm un}$ phase diagram showing mixed and demixed (phase separated) regions for different protein-solvent and inter-protein heterotypic interactions. In experimental studies, these interactions can be modulated either by altering the protein sequence (e.g., the ratio of hydrophilic to hydrophobic residues) or by tuning the properties of the solvent. As discussed in Sec.~\ref{secII}, here, we have taken a fixed protein concentration $\phi_{\rm p} = 0.5$ and chosen $J_{\rm uu} = 1.0$,  and $J_{\rm nn} = 0.5$. By altering the protein-solvent interactions, our model gives rise to different phase separation scenarios --- reentrant (Fig.~\ref{fig_2a}), UCST (Fig.~\ref{fig_2b}) and closed loop (Fig.~\ref{fig_2c}). For the reentrant phase scenario, we see that an entirely phase-separated state P1 appears at low temperature, where the proteins exist mostly in their native conformational state. An increase in $T$ results in the transition of this state to P2, where the thermal energy dominates over the weaker inter-(native) protein interaction, leaving the proteins in a homogeneously dispersed state. On further increase in $T$, the system transitions to a phase-separated state P3, in which the proteins are mostly in the unfolded state, and eventually we observe the dispersed phase (P4) at sufficiently high $T$ (see Fig.~\ref{fig_2a}). The representative snapshots and the corresponding protein composition profiles are shown in Figs.~\ref{fig_2d} and ~\ref{fig_2e}.  Thus, the phase diagram contains two distinct UCST lines --- designated as UCST-I and UCST-II --- situated between the phases P1-P2 and P3-P4, respectively, along with an LCST line between P2 and P3. This exhibits notable similarity with our previously reported results on implicit solvent lattice model~\cite{roy_anomalous_2025} as well as to previously reported instances of reentrant phase behavior observed in various biological systems (like IDPs and other biopolymers) characterized by presence of both UCST and LCST (for example, see Refs.~\citenum{ruff_advances_2018, dignon_temperature-controlled_2019}).

In Fig.~\ref{fig_2b}, while setting the protein-protein interaction parameters same as in Fig.~\ref{fig_2a}, we increased the unfolded-solvent interaction, $J_{\rm us}$, to $0.35$. It is evident from the figure that the P3 phase disappears from the phase diagram, leading to a complete UCST behavior, with a single UCST line present between the phases P1 and P2. We further probed another scenario where we have fixed the $J_{\rm us}$ corresponding to its value reported in Fig.~\ref{fig_2a} ($J_{\rm us}$ = 0.20) and increased the $J_{\rm ns}$ to $0.25$. In this scenario, we observe a closed loop phase diagram with LCST and UCST-II lines --- a phase diagram devoid of the P1 phase, with the notable difference of shifting of the LCST line to a slightly elevated temperature (see Fig.~\ref{fig_2c}). Also, for higher values of $J_{\rm un}$ ($\geq 0.36$), the system no longer exhibits phase-separated states, leaving only the homogeneous P2 and P4 phases, with the transition between them becoming continuous. To validate the robustness of our results on the choice of the initial condition, we explored: (i) demixed initial configuration with all proteins in native or unfolded states, and (ii) mixed initial configuration with completely random conformational distribution. The results show that the reported phase behavior remains invariant to modifications in the initial configurations. 

We further probed the effects of protein-solvent interactions on LLPS at non-equilibrium (driven) conditions, such non-Boltzmann protein conformational distribution among the two states and enhanced translational mobility of solvent and proteins due to the non-thermal (active) noise (see Sec.~I and Fig.~S1 in the Supplementary Materials). This study is motivated by the fact that the intracellular medium is inherently present in a non-equilibrium state, given the presence of different different active processes generating non thermal noise like ATP hydrolysis in molecular motors (myosin, kinesin, dynein)~\cite{chauhan_active_2025, mizuno_nonequilibrium_2007, fakhri_high-resolution_2014}. Our results suggest that these non-equilibrium conditions play a key role in shaping the rich phase behavior and, together with protein–solvent interactions, can serve as important regulatory parameters for controlling the LLPS of proteins.  
 \begin{figure*}[!t]
    \centering
    \begin{minipage}{\textwidth}
     \includegraphics[width=\textwidth]{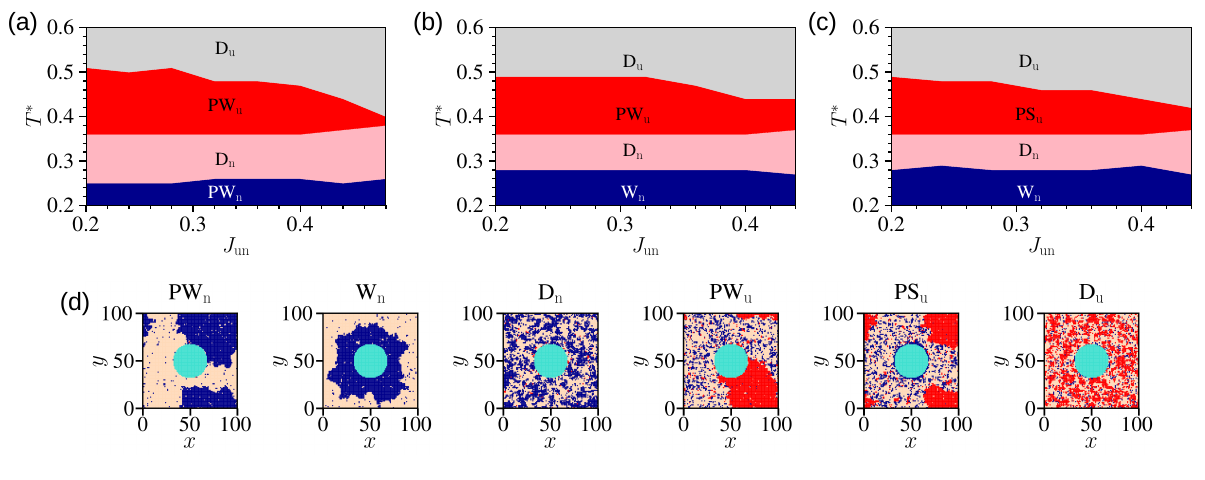}\label{fig4}
     \phantomsubcaption\label{fig_4a}
     \phantomsubcaption\label{fig_4b}
     \phantomsubcaption\label{fig_4c}
     \phantomsubcaption\label{fig_4d}
    \end{minipage}
    \caption{Reentrant phase behavior under different protein-cluster interactions is shown. Here, the total fraction of pinned particles is $0.1$, and $\phi_{\rm p}$ is kept fixed at $0.4$ such that the fraction of sites unoccupied by solvent is $0.5$ (same as in Fig.~\ref{fig_2a}). In all three cases, the unfolded protein-cluster interaction ($J_{\rm cu}$) is kept fixed at $0.6$, and the other model parameters are the same as in Fig.~\ref{fig_2a}. The native protein-cluster interaction ($J_{\rm cn}$) is set to $0.2$ (a), $0.6$ (b), and $1.0$ (c). The representative snapshots are shown in (d), where blue and red represent native and unfolded proteins respectively, while the impurity cluster is shown in cyan. Although the reentrant behavior remains broadly similar to that in Fig.~\ref{fig_2a}, the morphology of the various aggregated phases changes significantly with $J_{\rm cn}$. In (a), partial wetting of the impurity cluster by native and unfolded proteins (indicated by ${\rm PW_n}$ and ${\rm PW_u}$ in the snapshots in (d)) is observed in the low- and high-temperature phase-separated states, respectively. In (b), complete wetting of the cluster by the native protein occurs at low temperature (indicated by ${\rm W_n}$ in (d)), while the other phases remain similar to those in (a). Whereas, in (c), the unfolded proteins completely detach from the impurity cluster and form a segregated aggregate (indicated by ${\rm PS_u}$ in (d)). We further note that, in (b) and (c), the UCST line separating the ${\rm W_n}$ and ${\rm D_n}$ phases shifts toward higher temperatures compared to the corresponding line in (a), which separates the ${\rm PW_n}$ and ${\rm D_n}$ phases.}
\end{figure*}
\subsection{Origin of the solvent-induced alteration of LLPS}\label{sec_IIIB}
To gain a deeper understanding of how varying protein-solvent interactions alter the phase diagram, we first computed the temperature dependence of the conformational population --- measured as the average unfolded protein population, $\langle \phi_{\rm u} \rangle$ --- for three sets of values of the protein-solvent interaction parameter $J_{\rm ps}$ (where p = u and n). These three cases correspond to the distinct phase separation behaviors shown in Figs.~\ref{fig_2a}--\ref{fig_2c}. This analysis is motivated by the fact that the critical temperature (LCST or UCST) depends on the strength of inter-particle interactions, and therefore any alteration in the conformational population arising from conformation-dependent protein-solvent interactions will affect the dominant inter-particle interaction energy scale in the system, and thereby the overall phase behavior. We report that, at a fixed heterotypic interaction strength ($J_{\rm un} = 0.20$; see Fig.~\ref{fig_3a}), $\langle \phi_{\rm u} \rangle$ shows an abrupt jump as $T$ increases, with the transition being more pronounced for lower $J_{\rm ns}$ and higher $J_{\rm us}$. In this case, the mixing P1-to-P2 transition (see Fig.~\ref{fig_2a}; marked here by $\textsf{\textbf{+}}$) does not significantly alter the dominant protein population (here, native). However, upon the demixing transition (P2-to-P3, marked by $\times$), most native proteins abruptly transition to the unfolded phase separated state (P3), leading to a sharp increase in $\langle \phi_{\rm u} \rangle$. For higher heterotypic interaction strength $J_{\rm un}$, we observe a more gradual increase in the conformational population with increasing $T$, while the overall qualitative dependence on the protein-solvent interaction parameters ($J_{\rm ns}$ and $J_{\rm us}$) remains the same (see Fig.~S2 in the Supplementary Material).

To probe how protein-solvent interaction induced conformational population changes alter the mixed-demixed phase boundaries at different native and unfolded protein concentrations, in Fig.~\ref{fig_3b} we report the $T-\phi_{\rm p}$ phase diagram for a protein system consisting of purely native and purely unfolded states (no conformational change is allowed as $T$ increases) under varying protein-solvent interactions. We observe that the critical temperature (denoted as $T_{\rm cu}$ for unfolded proteins and $T_{\rm cn}$ for native proteins) is highly sensitive to protein-solvent interaction for both protein states and decreases with increasing protein-solvent interaction, leading to enhanced solubility or mixing of unfolded (native) proteins in the solvent as $J_{\rm us}$ ($J_{\rm ns}$) increases. Additionally, because unfolded proteins interact more strongly with each other than the native proteins (for a given protein-solvent interaction), the critical temperature for the unfolded state ($T_{\rm cu}$) is higher.

For the protein-solvent parameters belonging to the reentrant case (Fig.~\ref{fig_2a}), the low-$T$ P1-to-P2 transition corresponds to the system becoming supercritical with respect to $T_{\rm cn}$ (proteins are predominantly native). Upon further increasing $T$, the conformational population undergoes an abrupt transition to the unfolded state, raising the critical temperature to $T_{\rm cu}$. Consequently, the system becomes sub-critical with respect to $T_{\rm cu}$, leading to phase separation of unfolded proteins (P2-to-P3 transition). The system eventually becomes supercritical with respect to $T_{\rm cu}$ at sufficiently high $T$, and phase separation disappears. Increasing $J_{\rm us}$ to $0.35$ drastically reduces $T_{\rm cu}$, bringing it very close to $T_{\rm cn}$ (see Fig.~\ref{fig_3b}). As a result, once the native-dominated low-$T$ system becomes supercritical with respect to $T_{\rm cn}$, it almost simultaneously becomes supercritical with respect to $T_{\rm cu}$, producing a demixed phase throughout the higher-temperature region (see Fig.~\ref{fig_2b}) --- even though unfolded proteins interact more strongly among themselves than native proteins do. This gives rise to the exclusive UCST-type scenario reported in Fig.~\ref{fig_2b}. 

In the closed-loop phase diagram (Fig.~\ref{fig_2c}), a higher $J_{\rm ns}$ further lowers $T_{\rm cn}$ ($< 0.2$, not shown in Fig.~\ref{fig_3b}). Hence, the system remains supercritical with respect to $T_{\rm cn}$ even at low temperatures across the reported $T$ range. As $T$ increases, an abrupt conformational transition to the unfolded state occurs, promoting strong inter-unfolded protein interactions (the weak protein-solvent interaction also favors demixing). As a result, $T_{\rm cu}$ shifts to a higher value, making the system subcritical to this critical point and leading to a demixed phase (P2-to-P3 transition; reported in Fig.~\ref{fig_2c}). At even higher $T$, the system becomes supercritical again, returning to a homogeneous state and forming a closed-loop phase diagram. Thus, a delicate balance between conformation-dependent protein-solvent and protein-protein interactions shapes the phase behavior. Protein-solvent interactions --- tunable through protein sequence design or solvent properties --- can therefore be adjusted to engineer the phase diagram.  
\subsection{Effects of quenched (static) disorder on LLPS}\label{sec_IIIC}
Considering the degree of heterogeneity and crowding in the intracellular environment (like the presence of different macromolecular specifies and organelles), it becomes imperative to study the LLPS of proteins and other biomolecules in the presence of disorder of different length scales, rather than considering just a uniform one-component medium of proteins dispersed in solvent. Several factors, such as excluded volume and crowder-protein interactions,  can play key roles in phase separation kinetics and morphology. This section is primarily focused on understanding the effects of crowder-protein interaction on LLPS. To model crowders, we incorporated a cluster of immobile particles (i.e., static or quenched disorder) which interact differently with proteins in different conformation states and with solvents (see Sec.~\ref{secIIA}). We have considered here three scenarios: (i) unfolded proteins interact strongly with the frozen cluster particles than native ones, (ii) both native and unfolded proteins have similar interaction strength, and (iii) native proteins interact strongly with the cluster particles than unfolded ones. The total fraction of the particles contributing to static cluster is kept fixed at $0.1$, while the total protein fraction $\phi_{\rm p}$ is maintained at $0.4$, so that the fraction of sites unoccupied by solvent remain $0.5$ (same as in Fig.~\ref{fig2}). 

In Figs.~\ref{fig_4a}-\ref{fig_4c} we have reported the phase diagram for model parameters corresponding to the reentrant scenario (Fig.~\ref{fig_2a}) with three different protein-cluster interaction $J_{\rm cp}$ (with p = u and n) --- $J_{\rm cn}$ < $J_{\rm cu}$ (Fig.~\ref{fig_4a}),  $J_{\rm cn}$ $\sim$ $J_{\rm cu}$ (Fig.~\ref{fig_4b}) and $J_{\rm cn}$ > $J_{\rm cu}$ (Fig.~\ref{fig_4c}). We observe that the phase diagram remains broadly unchanged with respect to the case without the impurity cluster (see Fig.~\ref{fig_2a}), but with significantly different condensate morphologies, and therefore, we named differently the phases compared to Fig.~\ref{fig_2a}. For $J_{\rm cn}$ < $J_{\rm cu}$, the phase separated regions represent partial wetting of the impurity cluster by the native or unfolded proteins (referred to as PW$_{\rm n}$ and PW$_{\rm u}$, respectively; see Fig.~\ref{fig_4d}). However, for $J_{\rm cn}$ $\sim$ $J_{\rm cu}$, at low $T$, the impurity cluster is fully wetted by the native proteins throughout the range of the explored $J_{\rm un}$, with the PW$_{\rm u}$ phase appearing similar to the previous case. Lastly, for $J_{\rm cn} > J_{\rm cu}$, the PW$_{\rm n}$ phase is observed at low $T$, but for intermediate $T$ values, the cluster is no longer wetted by the unfolded proteins but a segregated phase of the same is seen (named as PS$_{\rm u}$, see Fig.~\ref{fig_2d}). Analogous to the P2 and P4 phases shown in Fig.~\ref{fig_2a}, the dispersed native and unfolded phases (D$_{\rm n}$ and D$_{\rm u}$) remains similar across Figs.~\ref{fig_4a}-\ref{fig_4c}. We further note that the UCST-I line (line separating PW$_{\rm n}$/W$_{\rm n}$ and D$_{\rm n}$ phases) is elevated to higher $T$ in for $J_{\rm cn}$ $\sim$ $J_{\rm cu}$ and $J_{\rm cn}$ > $J_{\rm cu}$. That is, the increase in $J_{\rm cn}$ (or introducing a new energy scale to the system) shifts the $T_{\rm cn}$ to the higher temperature and the system remains subcritical with respect to the same for a wider temperature range (see Fig.~S3 in the supplementary material). In the PS$_{\rm u}$ phase (Fig.~\ref{fig_4c} and the snapshot shown in Fig.~\ref{fig_4d}), the stronger interaction between native proteins and the cluster ``screens'' the latter, preventing the unfolded proteins from wetting its surface upon phase separation. 

We also separately studied the cases in which both $J_{\rm cn}$ and $J_{\rm cu}$ are comparable and low relative to $J_{\rm cs}$ (see Fig.~S4 in the supplementary material for details). They also show us that the condensate morphology is profoundly affected by the interplay of various interactions involved. In the case where $J_{\rm cn} = J_{\rm cu} =0.25$ (see Fig. S4(a)), the wetting of the impurity cluster by the unfolded proteins is entirely disrupted, resulting in the formation of a detached segregated agglomerate (the PS$_{\rm u}$ phase). A similar, albeit analogous effect occurs in the case of both native and unfolded proteins for a higher value of $J_{\rm cs}$ ($0.5$) (see Fig. S4(c) in supplementary material), where both the species form a detached segregative agglomerates (PS$_{\rm n}$ and PS$_{\rm u}$ phases) at low and intermediate $T$s, respectively. 
\begin{figure*}
    \centering
    \begin{minipage}{0.93\textwidth}
    \includegraphics[width=0.95\linewidth]{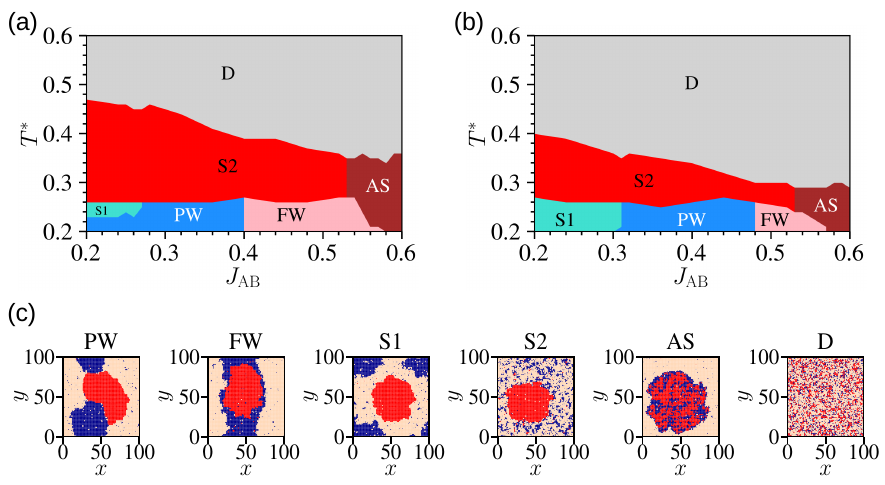}
    \phantomsubcaption\label{fig_5a}
    \phantomsubcaption\label{fig_5b}
    \phantomsubcaption\label{fig_5c}
    \end{minipage}
    \caption{Phase diagram showing the phase behavior of the binary protein mixture in the $T^*-J_{\rm AB}$ ($T^* = k_{\rm B}T/J_{\rm AA}$) plane, for two different protein-solvent interactions: $J_{\rm As} = 0.20$ (a) and $J_{\rm As} = 0.30$ (b). Note that $J_{\rm Bs} = 0.15$ in both (a) and (b). The observed condensate morphologies are: partially wetted (PW), fully wetted (FW), segregative (S1 and S2), associative (AS), and dispersed (D). Representative snapshots of the different condensate morphologies are shown in (c). Here, red, blue, and peach represent A protein, B protein, and solvent, respectively. In the PW phase, the weakly interacting B proteins are partly in contact with the strongly interacting A proteins. In contrast, in the FW phase, complete wetting of the A proteins by the B proteins is observed. The S1 phase occurs under conditions of low $T^*$ and low $J_{\rm AB}$, where both A and B proteins coexist in a segregated phase-separated state. The S2 phase, on the other hand, is characterized by A proteins residing in a segregated state while the B proteins remain dispersed in solvent. Comparing (a) and (b), we observe a shrinkage of the regions corresponding to the S2, AS, and FW phases, along with an enhancement of the S1 phase on increasing $J_{\rm As}$. A randomly mixed A:B protein composition ($50:50$) is used as the initial configuration for computing the phase diagrams.}
    \label{fig5}
\end{figure*}
\subsection{LLPS in binary protein mixture: Effects of protein-solvent interaction}\label{sec_IIID}
This part of work is motivated by recent advances in the area of multicomponent biomolecular phase organization with binary protein mixtures serving as tractable model systems for uncovering general thermodynamic principles~\cite{chew_thermodynamic_2023, wang_controlling_2025, rana_asymmetric_2024}. These studies highlight the importance of constituent states and their (homotypic and heterotypic) interactions in controlling binary miscibility. However, the role of protein–solvent interactions --- both in the absence and presence of crowders --- in determining the stability and morphology of multicomponent condensates remains poorly understood.

To elucidate the role of protein-solvent interaction in the shaping of condensate morphologies in binary protein mixtures, we mimicked the binary protein system through our lattice model. Here, the binary protein species (A and B) are represented by particles with same internal energy ($\epsilon = 0$). This condition is relevant for the protein mixture scenarios where the thermal energies are insufficient for the proteins to undergo major conformational changes. In  Figs.~\ref{fig_5a}-~\ref{fig_5b}, we show the $T-J_{\rm AB}$ phase diagram where we varied the heterotypic interaction $J_{\rm AB}$ keeping other parameters same as in Sec.~\ref{sec_IIIA} for two different protein-solvent interaction strengths --- $J_{\rm As} = 0.20$ (Fig.~\ref{fig_5a}) and $J_{\rm As} = 0.30$ (Fig.~\ref{fig_5b}). Here, we used a randomly mixed droplet of A and B proteins as the initial configuration to compute the phase diagram and reported the emergent equilibrium morphology. The total fraction of proteins in the system is kept fixed at $0.5$, with each type contributing an equal share of $0.25$, and A proteins have stronger interaction between themselves compared to B proteins. Similar to our implicit solvent model~\cite{roy_anomalous_2025}, this model yields a wide range of phase morphologies broadly classified as fully wetted, partially wetted, segregated, and associative. These morphologies are consistent with recent computational studies on model implicit-solvent binary protein systems, where the nature of inter-species interactions was shown to govern the rich condensate morphologies, including the formation of wetted, multilayered, and mixed condensates~\cite{chew_thermodynamic_2023, wang_controlling_2025, Ramachandran2026.03.04.709570}. 

It is evident from the figure that the protein-solvent interaction $J_{ \rm As}$ plays an important role in the stability of condensed morphology. For example, at low $J_{ \rm As}$ (see Fig.~\ref{fig_5a}), at low $T$ and low heterotypic interaction parameter values ($J_{\rm AB} \in [0.2, 0.4]$) we observe a partially wetted phase (PW) where some parts of the A-rich droplets are wetted by the B-rich phase (rest of the A-rich phase remains in direct contact with the solvent; see Fig.~\ref{fig_5c}. An increase in $J_{\rm As}$ (see Fig.~\ref{fig_5b}), at low $T$ makes the PW phase boundaries with other (S1, S2 and FW) transition towards higher $J_{\rm AB}$ ($J_{\rm AB} \in [0.31, 0.48]$). At low $J_{\rm AB}$ we observe the enhanced presence of the segregated S1 phase. Furthermore, the range of stability of the FW, S2 and AS phases in the $T^{*} -J_{\rm AB}$ plane also shrinks. The boundary separating the S2 and dispersed (D) phase shifts towards lower temperature. This boundary change can be attributed to the decrease in critical temperature corresponding to the A protein system as the protein-A-solvent interaction increases (see Section~\ref{sec_IIIB}). 
\begin{figure*}[t]
    \centering
    \begin{minipage}{\textwidth}
     \includegraphics[width=0.92\textwidth]{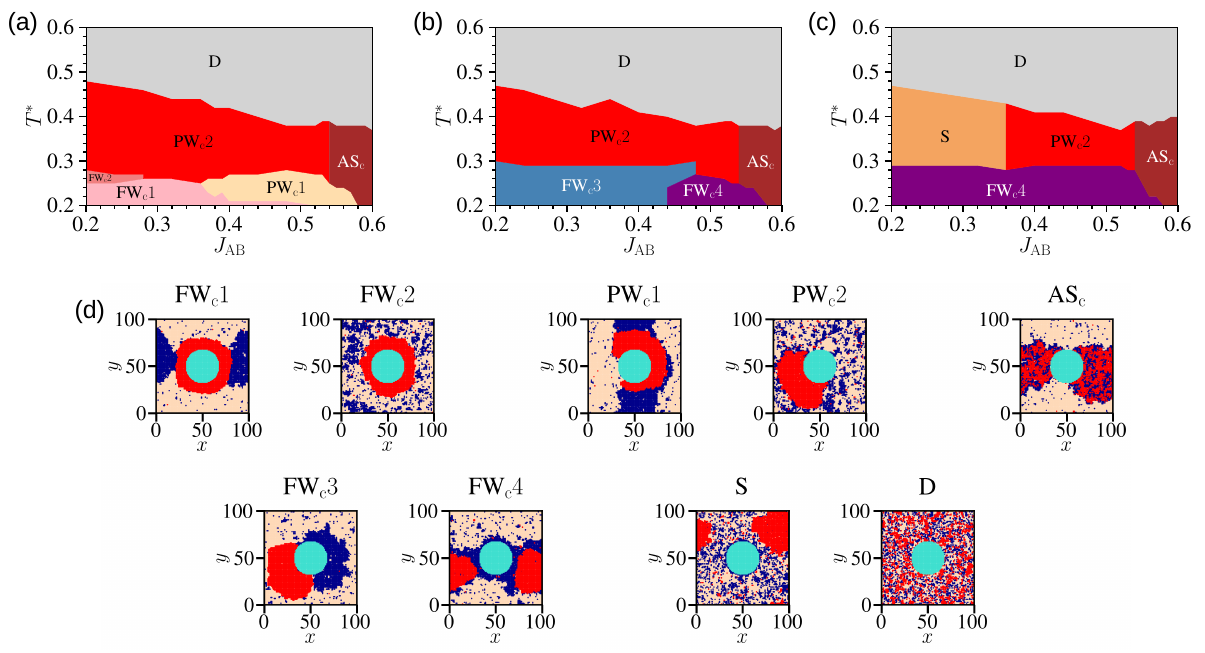}
     \phantomsubcaption\label{fig_6a}
     \phantomsubcaption\label{fig_6b}
     \phantomsubcaption\label{fig_6c}
     \phantomsubcaption\label{fig_6d}
    \end{minipage}
    \caption{Phase diagrams of the binary protein system in the presence of an impurity cluster (or crowder) for different protein-cluster interactions ($J_{\rm cp}$ with $p = \rm A, B$) in the $T^*-J_{\rm AB}$ plane. Here, $J_{\rm cA}$ is kept fixed at $0.6$, while $J_{\rm cB}$ is varied: $(J_{\rm cB} = 0.2$ (a), $0.6$ (b), and $1.0$ (c). All other model parameters are the same as those in Fig.~\ref{fig5}. Across (a)-(c), a drastic change in the wetting morphology of the cluster and protein condensate is observed in the low to intermediate temperature range. Representative snapshots of the different phases and condensate morphologies are shown in (d), where red and blue represent A and B proteins, respectively, and the impurity cluster is shown in cyan. In (a), at low temperature and low $J_{\rm AB}$, complete wetting of the cluster by A proteins is observed (${\rm FW_c1}$ and ${\rm FW_c2}$). In this regime, the weakly interacting B proteins partially wet the A proteins in the ${\rm FW_c1}$ phase and remain dispersed in the ${\rm FW_c2}$ phase. Increasing $J_{\rm AB}$ leads to a partially wetted phase (${\rm PW_c1}$), ultimately resulting in associative wetting of the cluster (${\rm AS_c}$). In (b) and (c), additional fully wetted condensate morphologies appear: ${\rm FW_c3}$ and ${\rm FW_c4}$ in (b), and ${\rm FW_c4}$ in (c). Other phases, such as ${\rm PW_c2}$ (partial wetting of the cluster by A proteins), segregative (S), and dispersed (D), are also observed.}
    \label{fig_6}
\end{figure*}
\subsection{LLPS in binary protein mixture: Effects of quenched disorder}  \label{sec_IIIE}
Finally, we have studied the effects of quenched disorder on the phase morphologies in the binary protein mixture system by incorporating an (static) impurity cluster (as described in Section~\ref{sec_IIIC}). We keep the A protein-cluster interaction fixed at $J_{\rm cA} = 0.6$ for all the cases probed here. On varying the B protein-cluster interaction, we computed $T - J_{\rm AB}$ phase diagram which shows a wide range of condensate morphologies in the low and intermediate $T$ range (see Figs.~\ref{fig_6a}-\ref{fig_6c}). The snapshots of the different condensate morphologies are shown in Fig.~\ref{fig_6d}. For $J_{\rm cB} < J_{\rm cA}$ (Fig.~\ref{fig_6a}), a full wetting of the cluster by A proteins which are further (partially) wetted by B proteins (named as FW$_{\rm c}1$; see Fig.~\ref{fig_6d}) is observed at low $T$ for low and intermediate $J_{\rm AB}$. This phase is taken over by the partially wetted cluster phase (PW$_{\rm c}1$) on increase in $J_{\rm AB}$, finally leading to the associative wetting of cluster (AS$_{\rm c}$) for higher $J_{\rm AB}$ ($\geq 0.54$). In the PW$_{\rm c}1$ phase, the cluster is partially wetted by the A proteins --- which are themselves wetted by the B proteins --- while the remaining part of the cluster surface remains exposed to the solvent. This phase emerges from the interplay between the relatively weaker cluster–B protein interaction, compared to the cluster-A protein interaction, and the comparatively stronger A-B interaction (although not sufficiently strong to stabilize the associative AS$_{\rm c}$ phase). As a result, the B proteins preferentially interacts and polarize the A proteins away from the cluster, leading to partial exposure of the cluster surface to the solvent. On increasing $T$ at low $J_{\rm AB}$ we get a narrow region of the FW$_{\rm c}2$ phase where the cluster is wetted by A proteins and B proteins are dispersed in the medium. On further increasing $T$ for $J_{\rm AB} \in $ [0.20, 0.54], FW$_{\rm c}1$, FW$_{\rm c}2$ and PW$_{\rm c}1$ transition to the PW$_{\rm c}2$ phase where the cluster is partially wetted by the A proteins and partially exposed to the solvent containing dispersed B proteins. At very low $J_{\rm AB} \sim 0.2$ and low $T$, the phase remains similar to FW$_{\rm c}$1 apart from one key difference: the B proteins do not come into direct contact with A proteins and instead form a segregative phase. 

For $J_{\rm cB} \sim J_{\rm cA}$ case (Fig.~\ref{fig_6b}), we observe a competitive wetting of the cluster by both A and B proteins (named as FW$_{\rm c}3$) phase in low $T$ and low $J_{\rm AB}$ region of the phase diagram. On increasing $J_{\rm AB}$, FW$_{\rm c}3$ phase transitions to FW$_{\rm c}4$ where the cluster is wetted by B proteins and A proteins wet the B protein condensate (opposite to FW$_{\rm c}1$ and PW$_{\rm c}1$ phases). Like the previous case, on increasing $T$ for $J_{\rm AB} \in $ [0.20, 0.54], low $T$ phase separated phases (FW$_{\rm c}3$ and FW$_{\rm c}4$) transition to the PW$_{\rm c}2$ phase, and we observe AS$_{\rm c}$ phase for $J_{\rm AB} > 0.54$. In the case of $J_{\rm cB} > J_{\rm cA}$ (Fig.~\ref{fig_6c}), the FW$_{\rm c}$4 phase remains the dominant phase at low $T$s (for $J_{\rm AB} \in $ [0.20, 0.54]), and transitions to AS$_{\rm c}$ at $J_{\rm AB} > 0.54$. However, a new segregative (S) phase of A proteins appear for $J_{\rm AB} \in [0.20, 0.36]$ at sufficiently high $T$s. The B type proteins remain dispersed uniformly in the solvent as well as around the impurity cluster. Since they interact strongly with the cluster (a thin layer of B proteins is observed on the cluster surface), the screening effect comes into play, which in turn prevents the wetting of the cluster by the A type proteins. This not only mimics the scenario where selective de-wetting is induced due to the dominance of a specific type of interaction over the others, but effectively captures the picture when selective adsorption of some species (in this case, B proteins) inhibits the binding of some other species. A detailed and extensive study focusing solely on effects of impurities (mimicking macromolecular crowding) and their spatial distribution, dynamics and interaction with solvent can shed more light on these aspects~\cite{andre_liquidliquid_2020, vweza_liquidliquid_2021, rauh_coarsegrained_2025}.
\section{Conclusion} \label{sec_IV}
We employed an explicit-solvent minimal statistical mechanical model based on the LG Hamiltonian with quenched disorder to investigate the roles of protein-solvent and protein-(static) disorder interactions (with static disorder mimicking intracellular crowders) on the phase behavior and morphology of biomolecular condensates. The proteins are represented as particles with two internal states: a non-degenerate ground state and a degenerate excited state representing two classes of protein conformational ensemble differing in their extent of folded (or unfolded) domains. The solvents are mimicked by non degenerate particles representing a volume element corresponding to a localized fluid volume of scale similar to the excluded volume of a protein unit. Additionally, the static disorder or crowder) is mimicked by cluster of pinned particles, which only interact with other species (like proteins and solvent). The computed model's phase diagram exhibits rich phase behavior, including UCST, closed-loop, and reentrant behavior under varying protein-solvent interaction conditions. We further elucidated the origin of the observed changes in phase behavior with varying protein-solvent interaction, and also investigated the role of static impurities in determining condensed phase morphology and phase stability.

Within the framework of our model, we also studied LLPS in binary protein mixtures in the presence and absence of static impurities to elucidate the roles of heterotypic protein-protein, protein-solvent, and protein-impurity interactions in governing phase behavior and condensed morphology. The binary mixture without crowder exhibits rich phase-separated morphologies, which can be broadly classified as partially wetted, fully wetted, segregative, and associative~\cite{chew_thermodynamic_2023, wang_controlling_2025, roy_anomalous_2025}. The phase boundaries between these morphologies are delicately sensitive to the protein-solvent interaction. Upon introducing disorder, the phase diagrams display a broad spectrum of complex condensate morphologies, determined by the choice of heterotypic protein-protein, protein-solvent, and protein-crowder interaction parameters.

Overall, our results demonstrate that the interplay among inter-protein (homotypic and heterotypic), protein-solvent, and protein-impurity (or crowder) interactions within the intracellular medium can generate a broad spectrum of mixed and demixed (phase-separated) phases in a multicomponent system. This work further underscores that protein-solvent and protein-crowder interactions represent key regulatory parameters for condensate morphology modulation. It is worth noting here is that extending our model to a system where the proteins have multiple internal states can be accomplished by implementing additional degrees of freedom, like in a $q$ state Potts model \cite{potts_generalized_1952, wu_potts_1982}. 

The sequence-dependent nature of inter-protein and protein-solvent interactions is intrinsic to proteins and biomolecular systems, and understanding their specificity in governing distinct phases is crucial to the study of LLPS. This, in turn, can guide future experimental and computational efforts in engineering and modulating specifically desired condensate morphologies. Future avenues of inquiry include the design of protein sequences that can mimic the inter-protein and protein-solvent interactions presented here, in order to test the predictions of our minimal model. Other potentially interesting directions would involve incorporating dynamical heterogeneities to probe their effects on phase behavior.

\begin{acknowledgments}
R.S.S. acknowledges financial support from ANRF (Grant No. CRG/2023/002975 and ANRF/ARGM/2025/002162/TS). S.R. acknowledges financial support from IISER Tirupati. The computations were performed at the IISER Tirupati computing facility.
\end{acknowledgments}

\bibliographystyle{aipnum4-1}
\bibliography{ts_llps_solvent} 

\begin{thebibliography}{96}%
\makeatletter
\providecommand \@ifxundefined [1]{%
 \@ifx{#1\undefined}
}%
\providecommand \@ifnum [1]{%
 \ifnum #1\expandafter \@firstoftwo
 \else \expandafter \@secondoftwo
 \fi
}%
\providecommand \@ifx [1]{%
 \ifx #1\expandafter \@firstoftwo
 \else \expandafter \@secondoftwo
 \fi
}%
\providecommand \natexlab [1]{#1}%
\providecommand \enquote  [1]{``#1''}%
\providecommand \bibnamefont  [1]{#1}%
\providecommand \bibfnamefont [1]{#1}%
\providecommand \citenamefont [1]{#1}%
\providecommand \href@noop [0]{\@secondoftwo}%
\providecommand \href [0]{\begingroup \@sanitize@url \@href}%
\providecommand \@href[1]{\@@startlink{#1}\@@href}%
\providecommand \@@href[1]{\endgroup#1\@@endlink}%
\providecommand \@sanitize@url [0]{\catcode `\\12\catcode `\$12\catcode
  `\&12\catcode `\#12\catcode `\^12\catcode `\_12\catcode `\%12\relax}%
\providecommand \@@startlink[1]{}%
\providecommand \@@endlink[0]{}%
\providecommand \url  [0]{\begingroup\@sanitize@url \@url }%
\providecommand \@url [1]{\endgroup\@href {#1}{\urlprefix }}%
\providecommand \urlprefix  [0]{URL }%
\providecommand \Eprint [0]{\href }%
\providecommand \doibase [0]{http://dx.doi.org/}%
\providecommand \selectlanguage [0]{\@gobble}%
\providecommand \bibinfo  [0]{\@secondoftwo}%
\providecommand \bibfield  [0]{\@secondoftwo}%
\providecommand \translation [1]{[#1]}%
\providecommand \BibitemOpen [0]{}%
\providecommand \bibitemStop [0]{}%
\providecommand \bibitemNoStop [0]{.\EOS\space}%
\providecommand \EOS [0]{\spacefactor3000\relax}%
\providecommand \BibitemShut  [1]{\csname bibitem#1\endcsname}%
\let\auto@bib@innerbib\@empty
\bibitem [{\citenamefont {Hyman}\ and\ \citenamefont
  {Simons}(2012)}]{hyman_beyond_2012}%
  \BibitemOpen
  \bibfield  {author} {\bibinfo {author} {\bibfnamefont {A.~A.}\ \bibnamefont
  {Hyman}}\ and\ \bibinfo {author} {\bibfnamefont {K.}~\bibnamefont {Simons}},\
  }\href {\doibase 10.1126/science.1223728} {\bibfield  {journal} {\bibinfo
  {journal} {Science}\ }\textbf {\bibinfo {volume} {337}},\ \bibinfo {pages}
  {1047} (\bibinfo {year} {2012})}\BibitemShut {NoStop}%
\bibitem [{\citenamefont {Hyman}, \citenamefont {Weber},\ and\ \citenamefont
  {Jülicher}(2014)}]{hyman_liquid-liquid_2014}%
  \BibitemOpen
  \bibfield  {author} {\bibinfo {author} {\bibfnamefont {A.~A.}\ \bibnamefont
  {Hyman}}, \bibinfo {author} {\bibfnamefont {C.~A.}\ \bibnamefont {Weber}}, \
  and\ \bibinfo {author} {\bibfnamefont {F.}~\bibnamefont {Jülicher}},\ }\href
  {\doibase 10.1146/annurev-cellbio-100913-013325} {\bibfield  {journal}
  {\bibinfo  {journal} {Annu. Rev. Cell Dev. Biol.}\ }\textbf {\bibinfo
  {volume} {30}},\ \bibinfo {pages} {39} (\bibinfo {year} {2014})}\BibitemShut
  {NoStop}%
\bibitem [{\citenamefont {Shin}\ and\ \citenamefont
  {Brangwynne}(2017)}]{shin_liquid_2017}%
  \BibitemOpen
  \bibfield  {author} {\bibinfo {author} {\bibfnamefont {Y.}~\bibnamefont
  {Shin}}\ and\ \bibinfo {author} {\bibfnamefont {C.~P.}\ \bibnamefont
  {Brangwynne}},\ }\href {\doibase 10.1126/science.aaf4382} {\bibfield
  {journal} {\bibinfo  {journal} {Science}\ }\textbf {\bibinfo {volume}
  {357}},\ \bibinfo {pages} {eaaf4382} (\bibinfo {year} {2017})}\BibitemShut
  {NoStop}%
\bibitem [{\citenamefont {Smokers}\ \emph {et~al.}(2024)\citenamefont
  {Smokers}, \citenamefont {Visser}, \citenamefont {Slootbeek}, \citenamefont
  {Huck},\ and\ \citenamefont {Spruijt}}]{smokers_how_2024}%
  \BibitemOpen
  \bibfield  {author} {\bibinfo {author} {\bibfnamefont {I.~B.~A.}\
  \bibnamefont {Smokers}}, \bibinfo {author} {\bibfnamefont {B.~S.}\
  \bibnamefont {Visser}}, \bibinfo {author} {\bibfnamefont {A.~D.}\
  \bibnamefont {Slootbeek}}, \bibinfo {author} {\bibfnamefont {W.~T.~S.}\
  \bibnamefont {Huck}}, \ and\ \bibinfo {author} {\bibfnamefont
  {E.}~\bibnamefont {Spruijt}},\ }\href {\doibase 10.1021/acs.accounts.4c00114}
  {\bibfield  {journal} {\bibinfo  {journal} {Acc. Chem. Res.}\ }\textbf
  {\bibinfo {volume} {57}},\ \bibinfo {pages} {1885} (\bibinfo {year}
  {2024})}\BibitemShut {NoStop}%
\bibitem [{\citenamefont {Das}\ \emph {et~al.}(2025)\citenamefont {Das},
  \citenamefont {Zaidi}, \citenamefont {Farag}, \citenamefont {Ruff},
  \citenamefont {Mahendran}, \citenamefont {Singh}, \citenamefont {Gui},
  \citenamefont {Messing}, \citenamefont {Taylor}, \citenamefont {Banerjee},
  \citenamefont {Pappu},\ and\ \citenamefont {Mittag}}]{das_tunable_2025}%
  \BibitemOpen
  \bibfield  {author} {\bibinfo {author} {\bibfnamefont {T.}~\bibnamefont
  {Das}}, \bibinfo {author} {\bibfnamefont {F.~K.}\ \bibnamefont {Zaidi}},
  \bibinfo {author} {\bibfnamefont {M.}~\bibnamefont {Farag}}, \bibinfo
  {author} {\bibfnamefont {K.~M.}\ \bibnamefont {Ruff}}, \bibinfo {author}
  {\bibfnamefont {T.~S.}\ \bibnamefont {Mahendran}}, \bibinfo {author}
  {\bibfnamefont {A.}~\bibnamefont {Singh}}, \bibinfo {author} {\bibfnamefont
  {X.}~\bibnamefont {Gui}}, \bibinfo {author} {\bibfnamefont {J.}~\bibnamefont
  {Messing}}, \bibinfo {author} {\bibfnamefont {J.~P.}\ \bibnamefont {Taylor}},
  \bibinfo {author} {\bibfnamefont {P.~R.}\ \bibnamefont {Banerjee}}, \bibinfo
  {author} {\bibfnamefont {R.~V.}\ \bibnamefont {Pappu}}, \ and\ \bibinfo
  {author} {\bibfnamefont {T.}~\bibnamefont {Mittag}},\ }\href {\doibase
  10.1016/j.molcel.2025.05.011} {\bibfield  {journal} {\bibinfo  {journal}
  {Mol. Cell}\ }\textbf {\bibinfo {volume} {85}},\ \bibinfo {pages} {2230}
  (\bibinfo {year} {2025})}\BibitemShut {NoStop}%
\bibitem [{\citenamefont {Peeples}\ and\ \citenamefont
  {Rosen}(2021)}]{peeples_mechanistic_2021}%
  \BibitemOpen
  \bibfield  {author} {\bibinfo {author} {\bibfnamefont {W.}~\bibnamefont
  {Peeples}}\ and\ \bibinfo {author} {\bibfnamefont {M.~K.}\ \bibnamefont
  {Rosen}},\ }\href {\doibase 10.1038/s41589-021-00801-x} {\bibfield  {journal}
  {\bibinfo  {journal} {Nat. Chem. Biol.}\ }\textbf {\bibinfo {volume} {17}},\
  \bibinfo {pages} {693} (\bibinfo {year} {2021})}\BibitemShut {NoStop}%
\bibitem [{\citenamefont {Riback}\ \emph {et~al.}(2017)\citenamefont {Riback},
  \citenamefont {Katanski}, \citenamefont {Kear-Scott}, \citenamefont
  {Pilipenko}, \citenamefont {Rojek}, \citenamefont {Sosnick},\ and\
  \citenamefont {Drummond}}]{riback_stress-triggered_2017}%
  \BibitemOpen
  \bibfield  {author} {\bibinfo {author} {\bibfnamefont {J.~A.}\ \bibnamefont
  {Riback}}, \bibinfo {author} {\bibfnamefont {C.~D.}\ \bibnamefont
  {Katanski}}, \bibinfo {author} {\bibfnamefont {J.~L.}\ \bibnamefont
  {Kear-Scott}}, \bibinfo {author} {\bibfnamefont {E.~V.}\ \bibnamefont
  {Pilipenko}}, \bibinfo {author} {\bibfnamefont {A.~E.}\ \bibnamefont
  {Rojek}}, \bibinfo {author} {\bibfnamefont {T.~R.}\ \bibnamefont {Sosnick}},
  \ and\ \bibinfo {author} {\bibfnamefont {D.~A.}\ \bibnamefont {Drummond}},\
  }\href {\doibase 10.1016/j.cell.2017.02.027} {\bibfield  {journal} {\bibinfo
  {journal} {Cell}\ }\textbf {\bibinfo {volume} {168}},\ \bibinfo {pages}
  {1028} (\bibinfo {year} {2017})}\BibitemShut {NoStop}%
\bibitem [{\citenamefont {Iserman}\ \emph {et~al.}(2020)\citenamefont
  {Iserman}, \citenamefont {Desroches~Altamirano}, \citenamefont {Jegers},
  \citenamefont {Friedrich}, \citenamefont {Zarin}, \citenamefont {Fritsch},
  \citenamefont {Mittasch}, \citenamefont {Domingues}, \citenamefont
  {Hersemann}, \citenamefont {Jahnel}, \citenamefont {Richter}, \citenamefont
  {Guenther}, \citenamefont {Hentze}, \citenamefont {Moses}, \citenamefont
  {Hyman}, \citenamefont {Kramer}, \citenamefont {Kreysing}, \citenamefont
  {Franzmann},\ and\ \citenamefont {Alberti}}]{iserman_condensation_2020}%
  \BibitemOpen
  \bibfield  {author} {\bibinfo {author} {\bibfnamefont {C.}~\bibnamefont
  {Iserman}}, \bibinfo {author} {\bibfnamefont {C.}~\bibnamefont
  {Desroches~Altamirano}}, \bibinfo {author} {\bibfnamefont {C.}~\bibnamefont
  {Jegers}}, \bibinfo {author} {\bibfnamefont {U.}~\bibnamefont {Friedrich}},
  \bibinfo {author} {\bibfnamefont {T.}~\bibnamefont {Zarin}}, \bibinfo
  {author} {\bibfnamefont {A.~W.}\ \bibnamefont {Fritsch}}, \bibinfo {author}
  {\bibfnamefont {M.}~\bibnamefont {Mittasch}}, \bibinfo {author}
  {\bibfnamefont {A.}~\bibnamefont {Domingues}}, \bibinfo {author}
  {\bibfnamefont {L.}~\bibnamefont {Hersemann}}, \bibinfo {author}
  {\bibfnamefont {M.}~\bibnamefont {Jahnel}}, \bibinfo {author} {\bibfnamefont
  {D.}~\bibnamefont {Richter}}, \bibinfo {author} {\bibfnamefont {U.-P.}\
  \bibnamefont {Guenther}}, \bibinfo {author} {\bibfnamefont {M.~W.}\
  \bibnamefont {Hentze}}, \bibinfo {author} {\bibfnamefont {A.~M.}\
  \bibnamefont {Moses}}, \bibinfo {author} {\bibfnamefont {A.~A.}\ \bibnamefont
  {Hyman}}, \bibinfo {author} {\bibfnamefont {G.}~\bibnamefont {Kramer}},
  \bibinfo {author} {\bibfnamefont {M.}~\bibnamefont {Kreysing}}, \bibinfo
  {author} {\bibfnamefont {T.~M.}\ \bibnamefont {Franzmann}}, \ and\ \bibinfo
  {author} {\bibfnamefont {S.}~\bibnamefont {Alberti}},\ }\href {\doibase
  10.1016/j.cell.2020.04.009} {\bibfield  {journal} {\bibinfo  {journal}
  {Cell}\ }\textbf {\bibinfo {volume} {181}},\ \bibinfo {pages} {818} (\bibinfo
  {year} {2020})}\BibitemShut {NoStop}%
\bibitem [{\citenamefont {Palumbo}\ \emph {et~al.}(2022)\citenamefont
  {Palumbo}, \citenamefont {McKean}, \citenamefont {Leatherman}, \citenamefont
  {Namitz}, \citenamefont {Connell}, \citenamefont {Wolfe}, \citenamefont
  {Moody}, \citenamefont {Gostinčar}, \citenamefont {Gunde-Cimerman},
  \citenamefont {Bah},\ and\ \citenamefont {Hanes}}]{palumbo_coevolution_2022}%
  \BibitemOpen
  \bibfield  {author} {\bibinfo {author} {\bibfnamefont {R.~J.}\ \bibnamefont
  {Palumbo}}, \bibinfo {author} {\bibfnamefont {N.}~\bibnamefont {McKean}},
  \bibinfo {author} {\bibfnamefont {E.}~\bibnamefont {Leatherman}}, \bibinfo
  {author} {\bibfnamefont {K.~E.~W.}\ \bibnamefont {Namitz}}, \bibinfo {author}
  {\bibfnamefont {L.}~\bibnamefont {Connell}}, \bibinfo {author} {\bibfnamefont
  {A.}~\bibnamefont {Wolfe}}, \bibinfo {author} {\bibfnamefont
  {K.}~\bibnamefont {Moody}}, \bibinfo {author} {\bibfnamefont
  {C.}~\bibnamefont {Gostinčar}}, \bibinfo {author} {\bibfnamefont
  {N.}~\bibnamefont {Gunde-Cimerman}}, \bibinfo {author} {\bibfnamefont
  {A.}~\bibnamefont {Bah}}, \ and\ \bibinfo {author} {\bibfnamefont {S.~D.}\
  \bibnamefont {Hanes}},\ }\href {\doibase 10.1126/sciadv.abq3235} {\bibfield
  {journal} {\bibinfo  {journal} {Sci. Adv.}\ }\textbf {\bibinfo {volume}
  {8}},\ \bibinfo {pages} {eabq3235} (\bibinfo {year} {2022})}\BibitemShut
  {NoStop}%
\bibitem [{\citenamefont {Alberti}\ \emph {et~al.}(2025)\citenamefont
  {Alberti}, \citenamefont {Arosio}, \citenamefont {Best}, \citenamefont
  {Boeynaems}, \citenamefont {Cai}, \citenamefont {Collepardo-Guevara},
  \citenamefont {Dignon}, \citenamefont {Dimova}, \citenamefont
  {Elbaum-Garfinkle}, \citenamefont {Fawzi}, \citenamefont {Fuxreiter},
  \citenamefont {Gladfelter}, \citenamefont {Honigmann}, \citenamefont {Jain},
  \citenamefont {Joseph}, \citenamefont {Knowles}, \citenamefont {Lasker},
  \citenamefont {Lemke}, \citenamefont {Lindorff-Larsen}, \citenamefont
  {Lipowsky}, \citenamefont {Mittal}, \citenamefont {Mukhopadhyay},
  \citenamefont {Myong}, \citenamefont {Pappu}, \citenamefont {Rippe},
  \citenamefont {Shelkovnikova}, \citenamefont {Vecchiarelli}, \citenamefont
  {Wegmann}, \citenamefont {Zhang}, \citenamefont {Zhang}, \citenamefont
  {Zubieta}, \citenamefont {Zweckstetter}, \citenamefont {Dormann},\ and\
  \citenamefont {Mittag}}]{alberti_current_2025}%
  \BibitemOpen
  \bibfield  {author} {\bibinfo {author} {\bibfnamefont {S.}~\bibnamefont
  {Alberti}}, \bibinfo {author} {\bibfnamefont {P.}~\bibnamefont {Arosio}},
  \bibinfo {author} {\bibfnamefont {R.~B.}\ \bibnamefont {Best}}, \bibinfo
  {author} {\bibfnamefont {S.}~\bibnamefont {Boeynaems}}, \bibinfo {author}
  {\bibfnamefont {D.}~\bibnamefont {Cai}}, \bibinfo {author} {\bibfnamefont
  {R.}~\bibnamefont {Collepardo-Guevara}}, \bibinfo {author} {\bibfnamefont
  {G.~L.}\ \bibnamefont {Dignon}}, \bibinfo {author} {\bibfnamefont
  {R.}~\bibnamefont {Dimova}}, \bibinfo {author} {\bibfnamefont
  {S.}~\bibnamefont {Elbaum-Garfinkle}}, \bibinfo {author} {\bibfnamefont
  {N.~L.}\ \bibnamefont {Fawzi}}, \bibinfo {author} {\bibfnamefont
  {M.}~\bibnamefont {Fuxreiter}}, \bibinfo {author} {\bibfnamefont {A.~S.}\
  \bibnamefont {Gladfelter}}, \bibinfo {author} {\bibfnamefont
  {A.}~\bibnamefont {Honigmann}}, \bibinfo {author} {\bibfnamefont
  {A.}~\bibnamefont {Jain}}, \bibinfo {author} {\bibfnamefont {J.~A.}\
  \bibnamefont {Joseph}}, \bibinfo {author} {\bibfnamefont {T.~P.~J.}\
  \bibnamefont {Knowles}}, \bibinfo {author} {\bibfnamefont {K.}~\bibnamefont
  {Lasker}}, \bibinfo {author} {\bibfnamefont {E.~A.}\ \bibnamefont {Lemke}},
  \bibinfo {author} {\bibfnamefont {K.}~\bibnamefont {Lindorff-Larsen}},
  \bibinfo {author} {\bibfnamefont {R.}~\bibnamefont {Lipowsky}}, \bibinfo
  {author} {\bibfnamefont {J.}~\bibnamefont {Mittal}}, \bibinfo {author}
  {\bibfnamefont {S.}~\bibnamefont {Mukhopadhyay}}, \bibinfo {author}
  {\bibfnamefont {S.}~\bibnamefont {Myong}}, \bibinfo {author} {\bibfnamefont
  {R.~V.}\ \bibnamefont {Pappu}}, \bibinfo {author} {\bibfnamefont
  {K.}~\bibnamefont {Rippe}}, \bibinfo {author} {\bibfnamefont {T.~A.}\
  \bibnamefont {Shelkovnikova}}, \bibinfo {author} {\bibfnamefont {A.~G.}\
  \bibnamefont {Vecchiarelli}}, \bibinfo {author} {\bibfnamefont
  {S.}~\bibnamefont {Wegmann}}, \bibinfo {author} {\bibfnamefont
  {H.}~\bibnamefont {Zhang}}, \bibinfo {author} {\bibfnamefont
  {M.}~\bibnamefont {Zhang}}, \bibinfo {author} {\bibfnamefont
  {C.}~\bibnamefont {Zubieta}}, \bibinfo {author} {\bibfnamefont
  {M.}~\bibnamefont {Zweckstetter}}, \bibinfo {author} {\bibfnamefont
  {D.}~\bibnamefont {Dormann}}, \ and\ \bibinfo {author} {\bibfnamefont
  {T.}~\bibnamefont {Mittag}},\ }\href {\doibase 10.1038/s41467-025-62055-8}
  {\bibfield  {journal} {\bibinfo  {journal} {Nat. Commun.}\ }\textbf {\bibinfo
  {volume} {16}},\ \bibinfo {pages} {7730} (\bibinfo {year}
  {2025})}\BibitemShut {NoStop}%
\bibitem [{\citenamefont {Wang}\ \emph {et~al.}(2021)\citenamefont {Wang},
  \citenamefont {Zhang}, \citenamefont {Dai}, \citenamefont {Lou},
  \citenamefont {Nie}, \citenamefont {Yan}, \citenamefont {Zhang},
  \citenamefont {Lan}, \citenamefont {Liu},\ and\ \citenamefont
  {Wang}}]{wang_liquidliquid_2021}%
  \BibitemOpen
  \bibfield  {author} {\bibinfo {author} {\bibfnamefont {B.}~\bibnamefont
  {Wang}}, \bibinfo {author} {\bibfnamefont {L.}~\bibnamefont {Zhang}},
  \bibinfo {author} {\bibfnamefont {T.}~\bibnamefont {Dai}}, \bibinfo {author}
  {\bibfnamefont {J.}~\bibnamefont {Lou}}, \bibinfo {author} {\bibfnamefont
  {J.}~\bibnamefont {Nie}}, \bibinfo {author} {\bibfnamefont {H.}~\bibnamefont
  {Yan}}, \bibinfo {author} {\bibfnamefont {Z.}~\bibnamefont {Zhang}}, \bibinfo
  {author} {\bibfnamefont {F.}~\bibnamefont {Lan}}, \bibinfo {author}
  {\bibfnamefont {J.}~\bibnamefont {Liu}}, \ and\ \bibinfo {author}
  {\bibfnamefont {L.}~\bibnamefont {Wang}},\ }\href {\doibase
  10.1038/s41392-021-00678-1} {\bibfield  {journal} {\bibinfo  {journal}
  {Signal Transduct. Target. Ther.}\ }\textbf {\bibinfo {volume} {6}},\
  \bibinfo {pages} {290} (\bibinfo {year} {2021})}\BibitemShut {NoStop}%
\bibitem [{\citenamefont {Alberti}\ and\ \citenamefont
  {Dormann}(2019)}]{alberti_liquidliquid_2019}%
  \BibitemOpen
  \bibfield  {author} {\bibinfo {author} {\bibfnamefont {S.}~\bibnamefont
  {Alberti}}\ and\ \bibinfo {author} {\bibfnamefont {D.}~\bibnamefont
  {Dormann}},\ }\href {\doibase 10.1146/annurev-genet-112618-043527} {\bibfield
   {journal} {\bibinfo  {journal} {Annu. Rev. Genet.}\ }\textbf {\bibinfo
  {volume} {53}},\ \bibinfo {pages} {171} (\bibinfo {year} {2019})}\BibitemShut
  {NoStop}%
\bibitem [{\citenamefont {Taylor}, \citenamefont {Hardy},\ and\ \citenamefont
  {Fischbeck}(2002)}]{taylor_toxic_2002}%
  \BibitemOpen
  \bibfield  {author} {\bibinfo {author} {\bibfnamefont {J.~P.}\ \bibnamefont
  {Taylor}}, \bibinfo {author} {\bibfnamefont {J.}~\bibnamefont {Hardy}}, \
  and\ \bibinfo {author} {\bibfnamefont {K.~H.}\ \bibnamefont {Fischbeck}},\
  }\href {\doibase 10.1126/science.1067122} {\bibfield  {journal} {\bibinfo
  {journal} {Science}\ }\textbf {\bibinfo {volume} {296}},\ \bibinfo {pages}
  {1991} (\bibinfo {year} {2002})}\BibitemShut {NoStop}%
\bibitem [{\citenamefont {Cai}, \citenamefont {Liu},\ and\ \citenamefont
  {Lippincott-Schwartz}(2021)}]{cai_biomolecular_2021}%
  \BibitemOpen
  \bibfield  {author} {\bibinfo {author} {\bibfnamefont {D.}~\bibnamefont
  {Cai}}, \bibinfo {author} {\bibfnamefont {Z.}~\bibnamefont {Liu}}, \ and\
  \bibinfo {author} {\bibfnamefont {J.}~\bibnamefont {Lippincott-Schwartz}},\
  }\href {\doibase 10.1016/j.tibs.2021.01.002} {\bibfield  {journal} {\bibinfo
  {journal} {Trends Biochem. Sci.}\ }\textbf {\bibinfo {volume} {46}},\
  \bibinfo {pages} {535} (\bibinfo {year} {2021})}\BibitemShut {NoStop}%
\bibitem [{\citenamefont {Song}\ \emph {et~al.}(2026)\citenamefont {Song},
  \citenamefont {Hao}, \citenamefont {Latacz}, \citenamefont {Cykowiak},
  \citenamefont {Kirylczuk}, \citenamefont {Quan}, \citenamefont {Palomba},
  \citenamefont {Ni}, \citenamefont {Liu}, \citenamefont {Hu}, \citenamefont
  {Shi}, \citenamefont {Posey}, \citenamefont {Li}, \citenamefont {Yuan},
  \citenamefont {Sun}, \citenamefont {Pappu}, \citenamefont {Digman},
  \citenamefont {Huang},\ and\ \citenamefont {Jiang}}]{song_unleashed_2026}%
  \BibitemOpen
  \bibfield  {author} {\bibinfo {author} {\bibfnamefont {Y.}~\bibnamefont
  {Song}}, \bibinfo {author} {\bibfnamefont {Y.}~\bibnamefont {Hao}}, \bibinfo
  {author} {\bibfnamefont {M.}~\bibnamefont {Latacz}}, \bibinfo {author}
  {\bibfnamefont {M.}~\bibnamefont {Cykowiak}}, \bibinfo {author}
  {\bibfnamefont {J.}~\bibnamefont {Kirylczuk}}, \bibinfo {author}
  {\bibfnamefont {X.}~\bibnamefont {Quan}}, \bibinfo {author} {\bibfnamefont
  {F.}~\bibnamefont {Palomba}}, \bibinfo {author} {\bibfnamefont
  {S.}~\bibnamefont {Ni}}, \bibinfo {author} {\bibfnamefont {L.}~\bibnamefont
  {Liu}}, \bibinfo {author} {\bibfnamefont {J.}~\bibnamefont {Hu}}, \bibinfo
  {author} {\bibfnamefont {B.}~\bibnamefont {Shi}}, \bibinfo {author}
  {\bibfnamefont {A.}~\bibnamefont {Posey}}, \bibinfo {author} {\bibfnamefont
  {Q.}~\bibnamefont {Li}}, \bibinfo {author} {\bibfnamefont {H.}~\bibnamefont
  {Yuan}}, \bibinfo {author} {\bibfnamefont {J.}~\bibnamefont {Sun}}, \bibinfo
  {author} {\bibfnamefont {R.}~\bibnamefont {Pappu}}, \bibinfo {author}
  {\bibfnamefont {M.~A.}\ \bibnamefont {Digman}}, \bibinfo {author}
  {\bibfnamefont {K.}~\bibnamefont {Huang}}, \ and\ \bibinfo {author}
  {\bibfnamefont {H.}~\bibnamefont {Jiang}},\ }\href {\doibase
  10.64898/2026.02.18.706652} {\bibfield  {journal} {\bibinfo  {journal}
  {bioRxiv}\ } (\bibinfo {year} {2026}),\
  10.64898/2026.02.18.706652}\BibitemShut {NoStop}%
\bibitem [{\citenamefont {McCormick}\ and\ \citenamefont
  {Khaperskyy}(2017)}]{mccormick_translation_2017}%
  \BibitemOpen
  \bibfield  {author} {\bibinfo {author} {\bibfnamefont {C.}~\bibnamefont
  {McCormick}}\ and\ \bibinfo {author} {\bibfnamefont {D.~A.}\ \bibnamefont
  {Khaperskyy}},\ }\href {\doibase 10.1038/nri.2017.63} {\bibfield  {journal}
  {\bibinfo  {journal} {Nat. Rev. Immunol.}\ }\textbf {\bibinfo {volume}
  {17}},\ \bibinfo {pages} {647} (\bibinfo {year} {2017})}\BibitemShut
  {NoStop}%
\bibitem [{\citenamefont {Savastano}\ \emph {et~al.}(2020)\citenamefont
  {Savastano}, \citenamefont {Ibáñez De~Opakua}, \citenamefont {Rankovic},\
  and\ \citenamefont {Zweckstetter}}]{savastano_nucleocapsid_2020}%
  \BibitemOpen
  \bibfield  {author} {\bibinfo {author} {\bibfnamefont {A.}~\bibnamefont
  {Savastano}}, \bibinfo {author} {\bibfnamefont {A.}~\bibnamefont {Ibáñez
  De~Opakua}}, \bibinfo {author} {\bibfnamefont {M.}~\bibnamefont {Rankovic}},
  \ and\ \bibinfo {author} {\bibfnamefont {M.}~\bibnamefont {Zweckstetter}},\
  }\href {\doibase 10.1038/s41467-020-19843-1} {\bibfield  {journal} {\bibinfo
  {journal} {Nat. Commun.}\ }\textbf {\bibinfo {volume} {11}},\ \bibinfo
  {pages} {6041} (\bibinfo {year} {2020})}\BibitemShut {NoStop}%
\bibitem [{\citenamefont {Cubuk}\ \emph {et~al.}(2021)\citenamefont {Cubuk},
  \citenamefont {Alston}, \citenamefont {Incicco}, \citenamefont {Singh},
  \citenamefont {Stuchell-Brereton}, \citenamefont {Ward}, \citenamefont
  {Zimmerman}, \citenamefont {Vithani}, \citenamefont {Griffith}, \citenamefont
  {Wagoner}, \citenamefont {Bowman}, \citenamefont {Hall}, \citenamefont
  {Soranno},\ and\ \citenamefont {Holehouse}}]{cubuk_sars-cov-2_2021}%
  \BibitemOpen
  \bibfield  {author} {\bibinfo {author} {\bibfnamefont {J.}~\bibnamefont
  {Cubuk}}, \bibinfo {author} {\bibfnamefont {J.~J.}\ \bibnamefont {Alston}},
  \bibinfo {author} {\bibfnamefont {J.~J.}\ \bibnamefont {Incicco}}, \bibinfo
  {author} {\bibfnamefont {S.}~\bibnamefont {Singh}}, \bibinfo {author}
  {\bibfnamefont {M.~D.}\ \bibnamefont {Stuchell-Brereton}}, \bibinfo {author}
  {\bibfnamefont {M.~D.}\ \bibnamefont {Ward}}, \bibinfo {author}
  {\bibfnamefont {M.~I.}\ \bibnamefont {Zimmerman}}, \bibinfo {author}
  {\bibfnamefont {N.}~\bibnamefont {Vithani}}, \bibinfo {author} {\bibfnamefont
  {D.}~\bibnamefont {Griffith}}, \bibinfo {author} {\bibfnamefont {J.~A.}\
  \bibnamefont {Wagoner}}, \bibinfo {author} {\bibfnamefont {G.~R.}\
  \bibnamefont {Bowman}}, \bibinfo {author} {\bibfnamefont {K.~B.}\
  \bibnamefont {Hall}}, \bibinfo {author} {\bibfnamefont {A.}~\bibnamefont
  {Soranno}}, \ and\ \bibinfo {author} {\bibfnamefont {A.~S.}\ \bibnamefont
  {Holehouse}},\ }\href {\doibase 10.1038/s41467-021-21953-3} {\bibfield
  {journal} {\bibinfo  {journal} {Nat. Commun.}\ }\textbf {\bibinfo {volume}
  {12}},\ \bibinfo {pages} {1936} (\bibinfo {year} {2021})}\BibitemShut
  {NoStop}%
\bibitem [{\citenamefont {Möller}\ \emph {et~al.}(2014)\citenamefont
  {Möller}, \citenamefont {Grobelny}, \citenamefont {Schulze}, \citenamefont
  {Bieder}, \citenamefont {Steffen}, \citenamefont {Erlkamp}, \citenamefont
  {Paulus}, \citenamefont {Tolan},\ and\ \citenamefont
  {Winter}}]{moller_reentrant_2014}%
  \BibitemOpen
  \bibfield  {author} {\bibinfo {author} {\bibfnamefont {J.}~\bibnamefont
  {Möller}}, \bibinfo {author} {\bibfnamefont {S.}~\bibnamefont {Grobelny}},
  \bibinfo {author} {\bibfnamefont {J.}~\bibnamefont {Schulze}}, \bibinfo
  {author} {\bibfnamefont {S.}~\bibnamefont {Bieder}}, \bibinfo {author}
  {\bibfnamefont {A.}~\bibnamefont {Steffen}}, \bibinfo {author} {\bibfnamefont
  {M.}~\bibnamefont {Erlkamp}}, \bibinfo {author} {\bibfnamefont
  {M.}~\bibnamefont {Paulus}}, \bibinfo {author} {\bibfnamefont
  {M.}~\bibnamefont {Tolan}}, \ and\ \bibinfo {author} {\bibfnamefont
  {R.}~\bibnamefont {Winter}},\ }\href {\doibase
  10.1103/physrevlett.112.028101} {\bibfield  {journal} {\bibinfo  {journal}
  {Phys. Rev. Lett.}\ }\textbf {\bibinfo {volume} {112}},\ \bibinfo {pages}
  {028101} (\bibinfo {year} {2014})}\BibitemShut {NoStop}%
\bibitem [{\citenamefont {Ruff}\ \emph {et~al.}(2018)\citenamefont {Ruff},
  \citenamefont {Roberts}, \citenamefont {Chilkoti},\ and\ \citenamefont
  {Pappu}}]{ruff_advances_2018}%
  \BibitemOpen
  \bibfield  {author} {\bibinfo {author} {\bibfnamefont {K.~M.}\ \bibnamefont
  {Ruff}}, \bibinfo {author} {\bibfnamefont {S.}~\bibnamefont {Roberts}},
  \bibinfo {author} {\bibfnamefont {A.}~\bibnamefont {Chilkoti}}, \ and\
  \bibinfo {author} {\bibfnamefont {R.~V.}\ \bibnamefont {Pappu}},\ }\href
  {\doibase 10.1016/j.jmb.2018.06.031} {\bibfield  {journal} {\bibinfo
  {journal} {J. Mol. Biol.}\ }\textbf {\bibinfo {volume} {430}},\ \bibinfo
  {pages} {4619} (\bibinfo {year} {2018})}\BibitemShut {NoStop}%
\bibitem [{\citenamefont {Tanaka}\ \emph {et~al.}(2024)\citenamefont {Tanaka},
  \citenamefont {Suyama}, \citenamefont {Tomohara},\ and\ \citenamefont
  {Nose}}]{tanaka_exploring_2024}%
  \BibitemOpen
  \bibfield  {author} {\bibinfo {author} {\bibfnamefont {N.}~\bibnamefont
  {Tanaka}}, \bibinfo {author} {\bibfnamefont {K.}~\bibnamefont {Suyama}},
  \bibinfo {author} {\bibfnamefont {K.}~\bibnamefont {Tomohara}}, \ and\
  \bibinfo {author} {\bibfnamefont {T.}~\bibnamefont {Nose}},\ }\href {\doibase
  10.1021/acs.biomac.4c00751} {\bibfield  {journal} {\bibinfo  {journal}
  {Biomacromolecules}\ }\textbf {\bibinfo {volume} {25}},\ \bibinfo {pages}
  {7156} (\bibinfo {year} {2024})}\BibitemShut {NoStop}%
\bibitem [{\citenamefont {Dignon}\ \emph {et~al.}(2019)\citenamefont {Dignon},
  \citenamefont {Zheng}, \citenamefont {Kim},\ and\ \citenamefont
  {Mittal}}]{dignon_temperature-controlled_2019}%
  \BibitemOpen
  \bibfield  {author} {\bibinfo {author} {\bibfnamefont {G.~L.}\ \bibnamefont
  {Dignon}}, \bibinfo {author} {\bibfnamefont {W.}~\bibnamefont {Zheng}},
  \bibinfo {author} {\bibfnamefont {Y.~C.}\ \bibnamefont {Kim}}, \ and\
  \bibinfo {author} {\bibfnamefont {J.}~\bibnamefont {Mittal}},\ }\href
  {\doibase 10.1021/acscentsci.9b00102} {\bibfield  {journal} {\bibinfo
  {journal} {ACS Cent. Sci.}\ }\textbf {\bibinfo {volume} {5}},\ \bibinfo
  {pages} {821} (\bibinfo {year} {2019})}\BibitemShut {NoStop}%
\bibitem [{\citenamefont {Adame-Arana}\ \emph {et~al.}(2020)\citenamefont
  {Adame-Arana}, \citenamefont {Weber}, \citenamefont {Zaburdaev},
  \citenamefont {Prost},\ and\ \citenamefont
  {Jülicher}}]{adame-arana_liquid_2020}%
  \BibitemOpen
  \bibfield  {author} {\bibinfo {author} {\bibfnamefont {O.}~\bibnamefont
  {Adame-Arana}}, \bibinfo {author} {\bibfnamefont {C.~A.}\ \bibnamefont
  {Weber}}, \bibinfo {author} {\bibfnamefont {V.}~\bibnamefont {Zaburdaev}},
  \bibinfo {author} {\bibfnamefont {J.}~\bibnamefont {Prost}}, \ and\ \bibinfo
  {author} {\bibfnamefont {F.}~\bibnamefont {Jülicher}},\ }\href {\doibase
  10.1016/j.bpj.2020.07.044} {\bibfield  {journal} {\bibinfo  {journal}
  {Biophys. J.}\ }\textbf {\bibinfo {volume} {119}},\ \bibinfo {pages} {1590}
  (\bibinfo {year} {2020})}\BibitemShut {NoStop}%
\bibitem [{\citenamefont {Cinar}\ \emph {et~al.}(2019)\citenamefont {Cinar},
  \citenamefont {Fetahaj}, \citenamefont {Cinar}, \citenamefont {Vernon},
  \citenamefont {Chan},\ and\ \citenamefont {Winter}}]{cinar_temperature_2019}%
  \BibitemOpen
  \bibfield  {author} {\bibinfo {author} {\bibfnamefont {H.}~\bibnamefont
  {Cinar}}, \bibinfo {author} {\bibfnamefont {Z.}~\bibnamefont {Fetahaj}},
  \bibinfo {author} {\bibfnamefont {S.}~\bibnamefont {Cinar}}, \bibinfo
  {author} {\bibfnamefont {R.~M.}\ \bibnamefont {Vernon}}, \bibinfo {author}
  {\bibfnamefont {H.~S.}\ \bibnamefont {Chan}}, \ and\ \bibinfo {author}
  {\bibfnamefont {R.~H.~A.}\ \bibnamefont {Winter}},\ }\href {\doibase
  10.1002/chem.201902210} {\bibfield  {journal} {\bibinfo  {journal} {Chem.
  Eur. J.}\ }\textbf {\bibinfo {volume} {25}},\ \bibinfo {pages} {13049}
  (\bibinfo {year} {2019})}\BibitemShut {NoStop}%
\bibitem [{\citenamefont {Li}\ \emph {et~al.}(2021)\citenamefont {Li},
  \citenamefont {Yoshizawa}, \citenamefont {Yamazaki}, \citenamefont
  {Fujiwara}, \citenamefont {Kameda},\ and\ \citenamefont
  {Kitahara}}]{li_pressure_2021}%
  \BibitemOpen
  \bibfield  {author} {\bibinfo {author} {\bibfnamefont {S.}~\bibnamefont
  {Li}}, \bibinfo {author} {\bibfnamefont {T.}~\bibnamefont {Yoshizawa}},
  \bibinfo {author} {\bibfnamefont {R.}~\bibnamefont {Yamazaki}}, \bibinfo
  {author} {\bibfnamefont {A.}~\bibnamefont {Fujiwara}}, \bibinfo {author}
  {\bibfnamefont {T.}~\bibnamefont {Kameda}}, \ and\ \bibinfo {author}
  {\bibfnamefont {R.}~\bibnamefont {Kitahara}},\ }\href {\doibase
  10.1021/acs.jpcb.1c01451} {\bibfield  {journal} {\bibinfo  {journal} {J.
  Phys. Chem. B}\ }\textbf {\bibinfo {volume} {125}},\ \bibinfo {pages} {6821}
  (\bibinfo {year} {2021})}\BibitemShut {NoStop}%
\bibitem [{\citenamefont {Cinar}\ \emph {et~al.}(2018)\citenamefont {Cinar},
  \citenamefont {Cinar}, \citenamefont {Chan},\ and\ \citenamefont
  {Winter}}]{cinar_pressureinduced_2018}%
  \BibitemOpen
  \bibfield  {author} {\bibinfo {author} {\bibfnamefont {H.}~\bibnamefont
  {Cinar}}, \bibinfo {author} {\bibfnamefont {S.}~\bibnamefont {Cinar}},
  \bibinfo {author} {\bibfnamefont {H.~S.}\ \bibnamefont {Chan}}, \ and\
  \bibinfo {author} {\bibfnamefont {R.}~\bibnamefont {Winter}},\ }\href
  {\doibase 10.1002/chem.201801643} {\bibfield  {journal} {\bibinfo  {journal}
  {Chem. Eur. J.}\ }\textbf {\bibinfo {volume} {24}},\ \bibinfo {pages} {8286}
  (\bibinfo {year} {2018})}\BibitemShut {NoStop}%
\bibitem [{\citenamefont {Joshi}\ \emph {et~al.}(2024)\citenamefont {Joshi},
  \citenamefont {Avni}, \citenamefont {Walimbe}, \citenamefont {Rai},
  \citenamefont {Sarkar},\ and\ \citenamefont
  {Mukhopadhyay}}]{joshi_hydrogen-bonded_2024}%
  \BibitemOpen
  \bibfield  {author} {\bibinfo {author} {\bibfnamefont {A.}~\bibnamefont
  {Joshi}}, \bibinfo {author} {\bibfnamefont {A.}~\bibnamefont {Avni}},
  \bibinfo {author} {\bibfnamefont {A.}~\bibnamefont {Walimbe}}, \bibinfo
  {author} {\bibfnamefont {S.~K.}\ \bibnamefont {Rai}}, \bibinfo {author}
  {\bibfnamefont {S.}~\bibnamefont {Sarkar}}, \ and\ \bibinfo {author}
  {\bibfnamefont {S.}~\bibnamefont {Mukhopadhyay}},\ }\href {\doibase
  10.1021/acs.jpclett.4c01153} {\bibfield  {journal} {\bibinfo  {journal} {J.
  Phys. Chem. Lett.}\ }\textbf {\bibinfo {volume} {15}},\ \bibinfo {pages}
  {7724} (\bibinfo {year} {2024})}\BibitemShut {NoStop}%
\bibitem [{\citenamefont {Krainer}\ \emph {et~al.}(2021)\citenamefont
  {Krainer}, \citenamefont {Welsh}, \citenamefont {Joseph}, \citenamefont
  {Espinosa}, \citenamefont {Wittmann}, \citenamefont {De~Csilléry},
  \citenamefont {Sridhar}, \citenamefont {Toprakcioglu}, \citenamefont
  {Gudiškytė}, \citenamefont {Czekalska}, \citenamefont {Arter},
  \citenamefont {Guillén-Boixet}, \citenamefont {Franzmann}, \citenamefont
  {Qamar}, \citenamefont {George-Hyslop}, \citenamefont {Hyman}, \citenamefont
  {Collepardo-Guevara}, \citenamefont {Alberti},\ and\ \citenamefont
  {Knowles}}]{krainer_reentrant_2021}%
  \BibitemOpen
  \bibfield  {author} {\bibinfo {author} {\bibfnamefont {G.}~\bibnamefont
  {Krainer}}, \bibinfo {author} {\bibfnamefont {T.~J.}\ \bibnamefont {Welsh}},
  \bibinfo {author} {\bibfnamefont {J.~A.}\ \bibnamefont {Joseph}}, \bibinfo
  {author} {\bibfnamefont {J.~R.}\ \bibnamefont {Espinosa}}, \bibinfo {author}
  {\bibfnamefont {S.}~\bibnamefont {Wittmann}}, \bibinfo {author}
  {\bibfnamefont {E.}~\bibnamefont {De~Csilléry}}, \bibinfo {author}
  {\bibfnamefont {A.}~\bibnamefont {Sridhar}}, \bibinfo {author} {\bibfnamefont
  {Z.}~\bibnamefont {Toprakcioglu}}, \bibinfo {author} {\bibfnamefont
  {G.}~\bibnamefont {Gudiškytė}}, \bibinfo {author} {\bibfnamefont {M.~A.}\
  \bibnamefont {Czekalska}}, \bibinfo {author} {\bibfnamefont {W.~E.}\
  \bibnamefont {Arter}}, \bibinfo {author} {\bibfnamefont {J.}~\bibnamefont
  {Guillén-Boixet}}, \bibinfo {author} {\bibfnamefont {T.~M.}\ \bibnamefont
  {Franzmann}}, \bibinfo {author} {\bibfnamefont {S.}~\bibnamefont {Qamar}},
  \bibinfo {author} {\bibfnamefont {P.~S.}\ \bibnamefont {George-Hyslop}},
  \bibinfo {author} {\bibfnamefont {A.~A.}\ \bibnamefont {Hyman}}, \bibinfo
  {author} {\bibfnamefont {R.}~\bibnamefont {Collepardo-Guevara}}, \bibinfo
  {author} {\bibfnamefont {S.}~\bibnamefont {Alberti}}, \ and\ \bibinfo
  {author} {\bibfnamefont {T.~P.~J.}\ \bibnamefont {Knowles}},\ }\href
  {\doibase 10.1038/s41467-021-21181-9} {\bibfield  {journal} {\bibinfo
  {journal} {Nat. Commun.}\ }\textbf {\bibinfo {volume} {12}},\ \bibinfo
  {pages} {1085} (\bibinfo {year} {2021})}\BibitemShut {NoStop}%
\bibitem [{\citenamefont {Quiroz}\ and\ \citenamefont
  {Chilkoti}(2015)}]{quiroz_sequence_2015}%
  \BibitemOpen
  \bibfield  {author} {\bibinfo {author} {\bibfnamefont {F.~G.}\ \bibnamefont
  {Quiroz}}\ and\ \bibinfo {author} {\bibfnamefont {A.}~\bibnamefont
  {Chilkoti}},\ }\href {\doibase 10.1038/nmat4418} {\bibfield  {journal}
  {\bibinfo  {journal} {Nat. Mater.}\ }\textbf {\bibinfo {volume} {14}},\
  \bibinfo {pages} {1164} (\bibinfo {year} {2015})}\BibitemShut {NoStop}%
\bibitem [{\citenamefont {Farag}\ \emph {et~al.}(2023)\citenamefont {Farag},
  \citenamefont {Borcherds}, \citenamefont {Bremer}, \citenamefont {Mittag},\
  and\ \citenamefont {Pappu}}]{farag_phase_2023}%
  \BibitemOpen
  \bibfield  {author} {\bibinfo {author} {\bibfnamefont {M.}~\bibnamefont
  {Farag}}, \bibinfo {author} {\bibfnamefont {W.~M.}\ \bibnamefont
  {Borcherds}}, \bibinfo {author} {\bibfnamefont {A.}~\bibnamefont {Bremer}},
  \bibinfo {author} {\bibfnamefont {T.}~\bibnamefont {Mittag}}, \ and\ \bibinfo
  {author} {\bibfnamefont {R.~V.}\ \bibnamefont {Pappu}},\ }\href {\doibase
  10.1038/s41467-023-41274-x} {\bibfield  {journal} {\bibinfo  {journal} {Nat.
  Commun.}\ }\textbf {\bibinfo {volume} {14}},\ \bibinfo {pages} {5527}
  (\bibinfo {year} {2023})}\BibitemShut {NoStop}%
\bibitem [{\citenamefont {Bremer}\ \emph {et~al.}(2022)\citenamefont {Bremer},
  \citenamefont {Farag}, \citenamefont {Borcherds}, \citenamefont {Peran},
  \citenamefont {Martin}, \citenamefont {Pappu},\ and\ \citenamefont
  {Mittag}}]{bremer_deciphering_2022}%
  \BibitemOpen
  \bibfield  {author} {\bibinfo {author} {\bibfnamefont {A.}~\bibnamefont
  {Bremer}}, \bibinfo {author} {\bibfnamefont {M.}~\bibnamefont {Farag}},
  \bibinfo {author} {\bibfnamefont {W.~M.}\ \bibnamefont {Borcherds}}, \bibinfo
  {author} {\bibfnamefont {I.}~\bibnamefont {Peran}}, \bibinfo {author}
  {\bibfnamefont {E.~W.}\ \bibnamefont {Martin}}, \bibinfo {author}
  {\bibfnamefont {R.~V.}\ \bibnamefont {Pappu}}, \ and\ \bibinfo {author}
  {\bibfnamefont {T.}~\bibnamefont {Mittag}},\ }\href {\doibase
  10.1038/s41557-021-00840-w} {\bibfield  {journal} {\bibinfo  {journal} {Nat.
  Chem.}\ }\textbf {\bibinfo {volume} {14}},\ \bibinfo {pages} {196} (\bibinfo
  {year} {2022})}\BibitemShut {NoStop}%
\bibitem [{\citenamefont {Martin}\ \emph {et~al.}(2020)\citenamefont {Martin},
  \citenamefont {Holehouse}, \citenamefont {Peran}, \citenamefont {Farag},
  \citenamefont {Incicco}, \citenamefont {Bremer}, \citenamefont {Grace},
  \citenamefont {Soranno}, \citenamefont {Pappu},\ and\ \citenamefont
  {Mittag}}]{martin_valence_2020}%
  \BibitemOpen
  \bibfield  {author} {\bibinfo {author} {\bibfnamefont {E.~W.}\ \bibnamefont
  {Martin}}, \bibinfo {author} {\bibfnamefont {A.~S.}\ \bibnamefont
  {Holehouse}}, \bibinfo {author} {\bibfnamefont {I.}~\bibnamefont {Peran}},
  \bibinfo {author} {\bibfnamefont {M.}~\bibnamefont {Farag}}, \bibinfo
  {author} {\bibfnamefont {J.~J.}\ \bibnamefont {Incicco}}, \bibinfo {author}
  {\bibfnamefont {A.}~\bibnamefont {Bremer}}, \bibinfo {author} {\bibfnamefont
  {C.~R.}\ \bibnamefont {Grace}}, \bibinfo {author} {\bibfnamefont
  {A.}~\bibnamefont {Soranno}}, \bibinfo {author} {\bibfnamefont {R.~V.}\
  \bibnamefont {Pappu}}, \ and\ \bibinfo {author} {\bibfnamefont
  {T.}~\bibnamefont {Mittag}},\ }\href {\doibase 10.1126/science.aaw8653}
  {\bibfield  {journal} {\bibinfo  {journal} {Science}\ }\textbf {\bibinfo
  {volume} {367}},\ \bibinfo {pages} {694} (\bibinfo {year}
  {2020})}\BibitemShut {NoStop}%
\bibitem [{\citenamefont {Dignon}\ \emph
  {et~al.}(2018{\natexlab{a}})\citenamefont {Dignon}, \citenamefont {Zheng},
  \citenamefont {Kim}, \citenamefont {Best},\ and\ \citenamefont
  {Mittal}}]{dignon_sequence_2018}%
  \BibitemOpen
  \bibfield  {author} {\bibinfo {author} {\bibfnamefont {G.~L.}\ \bibnamefont
  {Dignon}}, \bibinfo {author} {\bibfnamefont {W.}~\bibnamefont {Zheng}},
  \bibinfo {author} {\bibfnamefont {Y.~C.}\ \bibnamefont {Kim}}, \bibinfo
  {author} {\bibfnamefont {R.~B.}\ \bibnamefont {Best}}, \ and\ \bibinfo
  {author} {\bibfnamefont {J.}~\bibnamefont {Mittal}},\ }\href {\doibase
  10.1371/journal.pcbi.1005941} {\bibfield  {journal} {\bibinfo  {journal}
  {PLOS Comput. Biol.}\ }\textbf {\bibinfo {volume} {14}},\ \bibinfo {pages}
  {e1005941} (\bibinfo {year} {2018}{\natexlab{a}})}\BibitemShut {NoStop}%
\bibitem [{\citenamefont {Chew}\ \emph {et~al.}(2023)\citenamefont {Chew},
  \citenamefont {Joseph}, \citenamefont {Collepardo-Guevara},\ and\
  \citenamefont {Reinhardt}}]{chew_thermodynamic_2023}%
  \BibitemOpen
  \bibfield  {author} {\bibinfo {author} {\bibfnamefont {P.~Y.}\ \bibnamefont
  {Chew}}, \bibinfo {author} {\bibfnamefont {J.~A.}\ \bibnamefont {Joseph}},
  \bibinfo {author} {\bibfnamefont {R.}~\bibnamefont {Collepardo-Guevara}}, \
  and\ \bibinfo {author} {\bibfnamefont {A.}~\bibnamefont {Reinhardt}},\ }\href
  {\doibase 10.1039/D2SC05873A} {\bibfield  {journal} {\bibinfo  {journal}
  {Chem. Sci.}\ }\textbf {\bibinfo {volume} {14}},\ \bibinfo {pages} {1820}
  (\bibinfo {year} {2023})}\BibitemShut {NoStop}%
\bibitem [{\citenamefont {Rekhi}\ \emph {et~al.}(2024)\citenamefont {Rekhi},
  \citenamefont {Garcia}, \citenamefont {Barai}, \citenamefont {Rizuan},
  \citenamefont {Schuster}, \citenamefont {Kiick},\ and\ \citenamefont
  {Mittal}}]{rekhi_expanding_2024}%
  \BibitemOpen
  \bibfield  {author} {\bibinfo {author} {\bibfnamefont {S.}~\bibnamefont
  {Rekhi}}, \bibinfo {author} {\bibfnamefont {C.~G.}\ \bibnamefont {Garcia}},
  \bibinfo {author} {\bibfnamefont {M.}~\bibnamefont {Barai}}, \bibinfo
  {author} {\bibfnamefont {A.}~\bibnamefont {Rizuan}}, \bibinfo {author}
  {\bibfnamefont {B.~S.}\ \bibnamefont {Schuster}}, \bibinfo {author}
  {\bibfnamefont {K.~L.}\ \bibnamefont {Kiick}}, \ and\ \bibinfo {author}
  {\bibfnamefont {J.}~\bibnamefont {Mittal}},\ }\href {\doibase
  10.1038/s41557-024-01489-x} {\bibfield  {journal} {\bibinfo  {journal} {Nat.
  Chem.}\ }\textbf {\bibinfo {volume} {16}},\ \bibinfo {pages} {1113} (\bibinfo
  {year} {2024})}\BibitemShut {NoStop}%
\bibitem [{\citenamefont {Das}, \citenamefont {Ruff},\ and\ \citenamefont
  {Pappu}(2015)}]{das_relating_2015}%
  \BibitemOpen
  \bibfield  {author} {\bibinfo {author} {\bibfnamefont {R.~K.}\ \bibnamefont
  {Das}}, \bibinfo {author} {\bibfnamefont {K.~M.}\ \bibnamefont {Ruff}}, \
  and\ \bibinfo {author} {\bibfnamefont {R.~V.}\ \bibnamefont {Pappu}},\ }\href
  {\doibase 10.1016/j.sbi.2015.03.008} {\bibfield  {journal} {\bibinfo
  {journal} {Curr. Opin. Struct. Biol.}\ }\textbf {\bibinfo {volume} {32}},\
  \bibinfo {pages} {102} (\bibinfo {year} {2015})}\BibitemShut {NoStop}%
\bibitem [{\citenamefont {Benayad}\ \emph {et~al.}(2021)\citenamefont
  {Benayad}, \citenamefont {Von~Bülow}, \citenamefont {Stelzl},\ and\
  \citenamefont {Hummer}}]{benayad_simulation_2021}%
  \BibitemOpen
  \bibfield  {author} {\bibinfo {author} {\bibfnamefont {Z.}~\bibnamefont
  {Benayad}}, \bibinfo {author} {\bibfnamefont {S.}~\bibnamefont {Von~Bülow}},
  \bibinfo {author} {\bibfnamefont {L.~S.}\ \bibnamefont {Stelzl}}, \ and\
  \bibinfo {author} {\bibfnamefont {G.}~\bibnamefont {Hummer}},\ }\href
  {\doibase 10.1021/acs.jctc.0c01064} {\bibfield  {journal} {\bibinfo
  {journal} {J. Chem. Theory Comput.}\ }\textbf {\bibinfo {volume} {17}},\
  \bibinfo {pages} {525} (\bibinfo {year} {2021})}\BibitemShut {NoStop}%
\bibitem [{\citenamefont {Tesei}\ \emph {et~al.}(2021)\citenamefont {Tesei},
  \citenamefont {Schulze}, \citenamefont {Crehuet},\ and\ \citenamefont
  {Lindorff-Larsen}}]{tesei_accurate_2021}%
  \BibitemOpen
  \bibfield  {author} {\bibinfo {author} {\bibfnamefont {G.}~\bibnamefont
  {Tesei}}, \bibinfo {author} {\bibfnamefont {T.~K.}\ \bibnamefont {Schulze}},
  \bibinfo {author} {\bibfnamefont {R.}~\bibnamefont {Crehuet}}, \ and\
  \bibinfo {author} {\bibfnamefont {K.}~\bibnamefont {Lindorff-Larsen}},\
  }\href {\doibase 10.1073/pnas.2111696118} {\bibfield  {journal} {\bibinfo
  {journal} {Proc. Natl. Acad. Sci. U.S.A.}\ }\textbf {\bibinfo {volume}
  {118}},\ \bibinfo {pages} {e2111696118} (\bibinfo {year} {2021})}\BibitemShut
  {NoStop}%
\bibitem [{\citenamefont {Zeng}\ \emph {et~al.}(2021)\citenamefont {Zeng},
  \citenamefont {Liu}, \citenamefont {Fossat}, \citenamefont {Ren},
  \citenamefont {Chilkoti},\ and\ \citenamefont {Pappu}}]{zeng_design_2021}%
  \BibitemOpen
  \bibfield  {author} {\bibinfo {author} {\bibfnamefont {X.}~\bibnamefont
  {Zeng}}, \bibinfo {author} {\bibfnamefont {C.}~\bibnamefont {Liu}}, \bibinfo
  {author} {\bibfnamefont {M.~J.}\ \bibnamefont {Fossat}}, \bibinfo {author}
  {\bibfnamefont {P.}~\bibnamefont {Ren}}, \bibinfo {author} {\bibfnamefont
  {A.}~\bibnamefont {Chilkoti}}, \ and\ \bibinfo {author} {\bibfnamefont
  {R.~V.}\ \bibnamefont {Pappu}},\ }\href {\doibase 10.1063/5.0037438}
  {\bibfield  {journal} {\bibinfo  {journal} {APL Mater.}\ }\textbf {\bibinfo
  {volume} {9}},\ \bibinfo {pages} {021119} (\bibinfo {year}
  {2021})}\BibitemShut {NoStop}%
\bibitem [{\citenamefont {Simon}\ \emph {et~al.}(2017)\citenamefont {Simon},
  \citenamefont {Carroll}, \citenamefont {Rubinstein}, \citenamefont
  {Chilkoti},\ and\ \citenamefont {López}}]{simon_programming_2017}%
  \BibitemOpen
  \bibfield  {author} {\bibinfo {author} {\bibfnamefont {J.~R.}\ \bibnamefont
  {Simon}}, \bibinfo {author} {\bibfnamefont {N.~J.}\ \bibnamefont {Carroll}},
  \bibinfo {author} {\bibfnamefont {M.}~\bibnamefont {Rubinstein}}, \bibinfo
  {author} {\bibfnamefont {A.}~\bibnamefont {Chilkoti}}, \ and\ \bibinfo
  {author} {\bibfnamefont {G.~P.}\ \bibnamefont {López}},\ }\href {\doibase
  10.1038/nchem.2715} {\bibfield  {journal} {\bibinfo  {journal} {Nat. Chem.}\
  }\textbf {\bibinfo {volume} {9}},\ \bibinfo {pages} {509} (\bibinfo {year}
  {2017})}\BibitemShut {NoStop}%
\bibitem [{\citenamefont {Chang}\ \emph {et~al.}(2017)\citenamefont {Chang},
  \citenamefont {Lytle}, \citenamefont {Radhakrishna}, \citenamefont {Madinya},
  \citenamefont {Vélez}, \citenamefont {Sing},\ and\ \citenamefont
  {Perry}}]{chang_sequence_2017}%
  \BibitemOpen
  \bibfield  {author} {\bibinfo {author} {\bibfnamefont {L.-W.}\ \bibnamefont
  {Chang}}, \bibinfo {author} {\bibfnamefont {T.~K.}\ \bibnamefont {Lytle}},
  \bibinfo {author} {\bibfnamefont {M.}~\bibnamefont {Radhakrishna}}, \bibinfo
  {author} {\bibfnamefont {J.~J.}\ \bibnamefont {Madinya}}, \bibinfo {author}
  {\bibfnamefont {J.}~\bibnamefont {Vélez}}, \bibinfo {author} {\bibfnamefont
  {C.~E.}\ \bibnamefont {Sing}}, \ and\ \bibinfo {author} {\bibfnamefont
  {S.~L.}\ \bibnamefont {Perry}},\ }\href {\doibase 10.1038/s41467-017-01249-1}
  {\bibfield  {journal} {\bibinfo  {journal} {Nat. Commun.}\ }\textbf {\bibinfo
  {volume} {8}},\ \bibinfo {pages} {1273} (\bibinfo {year} {2017})}\BibitemShut
  {NoStop}%
\bibitem [{\citenamefont {André}\ and\ \citenamefont
  {Spruijt}(2020)}]{andre_liquidliquid_2020}%
  \BibitemOpen
  \bibfield  {author} {\bibinfo {author} {\bibfnamefont {A.~A.~M.}\
  \bibnamefont {André}}\ and\ \bibinfo {author} {\bibfnamefont
  {E.}~\bibnamefont {Spruijt}},\ }\href {\doibase 10.3390/ijms21165908}
  {\bibfield  {journal} {\bibinfo  {journal} {Int. J. Mol. Sci.}\ }\textbf
  {\bibinfo {volume} {21}},\ \bibinfo {pages} {5908} (\bibinfo {year}
  {2020})}\BibitemShut {NoStop}%
\bibitem [{\citenamefont {Vweza}, \citenamefont {Song},\ and\ \citenamefont
  {Chong}(2021)}]{vweza_liquidliquid_2021}%
  \BibitemOpen
  \bibfield  {author} {\bibinfo {author} {\bibfnamefont {A.-O.}\ \bibnamefont
  {Vweza}}, \bibinfo {author} {\bibfnamefont {C.-G.}\ \bibnamefont {Song}}, \
  and\ \bibinfo {author} {\bibfnamefont {K.-T.}\ \bibnamefont {Chong}},\ }\href
  {\doibase 10.3390/ijms22136675} {\bibfield  {journal} {\bibinfo  {journal}
  {Int. J. Mol. Sci.}\ }\textbf {\bibinfo {volume} {22}},\ \bibinfo {pages}
  {6675} (\bibinfo {year} {2021})}\BibitemShut {NoStop}%
\bibitem [{\citenamefont {Muthukumar}\ \emph {et~al.}(2026)\citenamefont
  {Muthukumar}, \citenamefont {Sundaravadivelu~Devarajan}, \citenamefont
  {Kim},\ and\ \citenamefont {Mittal}}]{muthukumar_sticky_2026}%
  \BibitemOpen
  \bibfield  {author} {\bibinfo {author} {\bibfnamefont {K.}~\bibnamefont
  {Muthukumar}}, \bibinfo {author} {\bibfnamefont {D.}~\bibnamefont
  {Sundaravadivelu~Devarajan}}, \bibinfo {author} {\bibfnamefont {Y.~C.}\
  \bibnamefont {Kim}}, \ and\ \bibinfo {author} {\bibfnamefont
  {J.}~\bibnamefont {Mittal}},\ }\href {\doibase 10.1073/pnas.2518384122}
  {\bibfield  {journal} {\bibinfo  {journal} {Proc. Natl. Acad. Sci. U.S.A.}\
  }\textbf {\bibinfo {volume} {123}},\ \bibinfo {pages} {e2518384122} (\bibinfo
  {year} {2026})}\BibitemShut {NoStop}%
\bibitem [{\citenamefont {Ausserwöger}\ \emph {et~al.}(2025)\citenamefont
  {Ausserwöger}, \citenamefont {De~Csilléry}, \citenamefont {Qian},
  \citenamefont {Krainer}, \citenamefont {Welsh}, \citenamefont {Sneideris},
  \citenamefont {Franzmann}, \citenamefont {Qamar}, \citenamefont {Erkamp},
  \citenamefont {Nixon-Abell}, \citenamefont {Kar}, \citenamefont
  {St~George-Hyslop}, \citenamefont {Hyman}, \citenamefont {Alberti},
  \citenamefont {Pappu},\ and\ \citenamefont
  {Knowles}}]{ausserwoger_quantifying_2025}%
  \BibitemOpen
  \bibfield  {author} {\bibinfo {author} {\bibfnamefont {H.}~\bibnamefont
  {Ausserwöger}}, \bibinfo {author} {\bibfnamefont {E.}~\bibnamefont
  {De~Csilléry}}, \bibinfo {author} {\bibfnamefont {D.}~\bibnamefont {Qian}},
  \bibinfo {author} {\bibfnamefont {G.}~\bibnamefont {Krainer}}, \bibinfo
  {author} {\bibfnamefont {T.~J.}\ \bibnamefont {Welsh}}, \bibinfo {author}
  {\bibfnamefont {T.}~\bibnamefont {Sneideris}}, \bibinfo {author}
  {\bibfnamefont {T.~M.}\ \bibnamefont {Franzmann}}, \bibinfo {author}
  {\bibfnamefont {S.}~\bibnamefont {Qamar}}, \bibinfo {author} {\bibfnamefont
  {N.~A.}\ \bibnamefont {Erkamp}}, \bibinfo {author} {\bibfnamefont
  {J.}~\bibnamefont {Nixon-Abell}}, \bibinfo {author} {\bibfnamefont
  {M.}~\bibnamefont {Kar}}, \bibinfo {author} {\bibfnamefont {P.}~\bibnamefont
  {St~George-Hyslop}}, \bibinfo {author} {\bibfnamefont {A.~A.}\ \bibnamefont
  {Hyman}}, \bibinfo {author} {\bibfnamefont {S.}~\bibnamefont {Alberti}},
  \bibinfo {author} {\bibfnamefont {R.~V.}\ \bibnamefont {Pappu}}, \ and\
  \bibinfo {author} {\bibfnamefont {T.~P.~J.}\ \bibnamefont {Knowles}},\ }\href
  {https://www.nature.com/articles/s41467-025-62437-y} {\bibfield  {journal}
  {\bibinfo  {journal} {Nat. Commun.}\ }\textbf {\bibinfo {volume} {16}},\
  \bibinfo {pages} {7724} (\bibinfo {year} {2025})}\BibitemShut {NoStop}%
\bibitem [{\citenamefont {Von~Bülow}\ \emph {et~al.}(2025)\citenamefont
  {Von~Bülow}, \citenamefont {Tesei}, \citenamefont {Zaidi}, \citenamefont
  {Mittag},\ and\ \citenamefont {Lindorff-Larsen}}]{von_bulow_prediction_2025}%
  \BibitemOpen
  \bibfield  {author} {\bibinfo {author} {\bibfnamefont {S.}~\bibnamefont
  {Von~Bülow}}, \bibinfo {author} {\bibfnamefont {G.}~\bibnamefont {Tesei}},
  \bibinfo {author} {\bibfnamefont {F.~K.}\ \bibnamefont {Zaidi}}, \bibinfo
  {author} {\bibfnamefont {T.}~\bibnamefont {Mittag}}, \ and\ \bibinfo {author}
  {\bibfnamefont {K.}~\bibnamefont {Lindorff-Larsen}},\ }\href {\doibase
  10.1073/pnas.2417920122} {\bibfield  {journal} {\bibinfo  {journal} {Proc.
  Natl. Acad. Sci. U.S.A.}\ }\textbf {\bibinfo {volume} {122}},\ \bibinfo
  {pages} {e2417920122} (\bibinfo {year} {2025})}\BibitemShut {NoStop}%
\bibitem [{\citenamefont {Feric}\ \emph {et~al.}(2016)\citenamefont {Feric},
  \citenamefont {Vaidya}, \citenamefont {Harmon}, \citenamefont {Mitrea},
  \citenamefont {Zhu}, \citenamefont {Richardson}, \citenamefont {Kriwacki},
  \citenamefont {Pappu},\ and\ \citenamefont
  {Brangwynne}}]{feric_coexisting_2016}%
  \BibitemOpen
  \bibfield  {author} {\bibinfo {author} {\bibfnamefont {M.}~\bibnamefont
  {Feric}}, \bibinfo {author} {\bibfnamefont {N.}~\bibnamefont {Vaidya}},
  \bibinfo {author} {\bibfnamefont {T.~S.}\ \bibnamefont {Harmon}}, \bibinfo
  {author} {\bibfnamefont {D.~M.}\ \bibnamefont {Mitrea}}, \bibinfo {author}
  {\bibfnamefont {L.}~\bibnamefont {Zhu}}, \bibinfo {author} {\bibfnamefont
  {T.~M.}\ \bibnamefont {Richardson}}, \bibinfo {author} {\bibfnamefont
  {R.~W.}\ \bibnamefont {Kriwacki}}, \bibinfo {author} {\bibfnamefont {R.~V.}\
  \bibnamefont {Pappu}}, \ and\ \bibinfo {author} {\bibfnamefont {C.~P.}\
  \bibnamefont {Brangwynne}},\ }\href {\doibase 10.1016/j.cell.2016.04.047}
  {\bibfield  {journal} {\bibinfo  {journal} {Cell}\ }\textbf {\bibinfo
  {volume} {165}},\ \bibinfo {pages} {1686} (\bibinfo {year}
  {2016})}\BibitemShut {NoStop}%
\bibitem [{\citenamefont {Fare}\ \emph {et~al.}(2021)\citenamefont {Fare},
  \citenamefont {Villani}, \citenamefont {Drake},\ and\ \citenamefont
  {Shorter}}]{fare_higher-order_2021}%
  \BibitemOpen
  \bibfield  {author} {\bibinfo {author} {\bibfnamefont {C.~M.}\ \bibnamefont
  {Fare}}, \bibinfo {author} {\bibfnamefont {A.}~\bibnamefont {Villani}},
  \bibinfo {author} {\bibfnamefont {L.~E.}\ \bibnamefont {Drake}}, \ and\
  \bibinfo {author} {\bibfnamefont {J.}~\bibnamefont {Shorter}},\ }\href
  {\doibase 10.1098/rsob.210137} {\bibfield  {journal} {\bibinfo  {journal}
  {Open Biol.}\ }\textbf {\bibinfo {volume} {11}},\ \bibinfo {pages} {210137}
  (\bibinfo {year} {2021})}\BibitemShut {NoStop}%
\bibitem [{\citenamefont {Regy}\ \emph {et~al.}(2020)\citenamefont {Regy},
  \citenamefont {Dignon}, \citenamefont {Zheng}, \citenamefont {Kim},\ and\
  \citenamefont {Mittal}}]{regy_sequence_2020}%
  \BibitemOpen
  \bibfield  {author} {\bibinfo {author} {\bibfnamefont {R.~M.}\ \bibnamefont
  {Regy}}, \bibinfo {author} {\bibfnamefont {G.~L.}\ \bibnamefont {Dignon}},
  \bibinfo {author} {\bibfnamefont {W.}~\bibnamefont {Zheng}}, \bibinfo
  {author} {\bibfnamefont {Y.~C.}\ \bibnamefont {Kim}}, \ and\ \bibinfo
  {author} {\bibfnamefont {J.}~\bibnamefont {Mittal}},\ }\href {\doibase
  10.1093/nar/gkaa1099} {\bibfield  {journal} {\bibinfo  {journal} {Nucleic
  Acids Res.}\ }\textbf {\bibinfo {volume} {48}},\ \bibinfo {pages} {12593}
  (\bibinfo {year} {2020})}\BibitemShut {NoStop}%
\bibitem [{\citenamefont {Kelley}\ \emph {et~al.}(2021)\citenamefont {Kelley},
  \citenamefont {Favetta}, \citenamefont {Regy}, \citenamefont {Mittal},\ and\
  \citenamefont {Schuster}}]{kelley_amphiphilic_2021}%
  \BibitemOpen
  \bibfield  {author} {\bibinfo {author} {\bibfnamefont {F.~M.}\ \bibnamefont
  {Kelley}}, \bibinfo {author} {\bibfnamefont {B.}~\bibnamefont {Favetta}},
  \bibinfo {author} {\bibfnamefont {R.~M.}\ \bibnamefont {Regy}}, \bibinfo
  {author} {\bibfnamefont {J.}~\bibnamefont {Mittal}}, \ and\ \bibinfo {author}
  {\bibfnamefont {B.~S.}\ \bibnamefont {Schuster}},\ }\href {\doibase
  10.1073/pnas.2109967118} {\bibfield  {journal} {\bibinfo  {journal} {Proc.
  Natl. Acad. Sci. U.S.A.}\ }\textbf {\bibinfo {volume} {118}},\ \bibinfo
  {pages} {e2109967118} (\bibinfo {year} {2021})}\BibitemShut {NoStop}%
\bibitem [{\citenamefont {Jiang}\ and\ \citenamefont
  {Ha-Duong}(2025)}]{jiang_temperature-dependent_2025}%
  \BibitemOpen
  \bibfield  {author} {\bibinfo {author} {\bibfnamefont {Y.}~\bibnamefont
  {Jiang}}\ and\ \bibinfo {author} {\bibfnamefont {T.}~\bibnamefont
  {Ha-Duong}},\ }\href {\doibase 10.1021/acs.jctc.5c00212} {\bibfield
  {journal} {\bibinfo  {journal} {J. Chem. Theory Comput.}\ }\textbf {\bibinfo
  {volume} {21}},\ \bibinfo {pages} {4939} (\bibinfo {year}
  {2025})}\BibitemShut {NoStop}%
\bibitem [{\citenamefont {Dhamankar}\ and\ \citenamefont
  {Webb}(2024)}]{dhamankar_asymmetry_2024}%
  \BibitemOpen
  \bibfield  {author} {\bibinfo {author} {\bibfnamefont {S.}~\bibnamefont
  {Dhamankar}}\ and\ \bibinfo {author} {\bibfnamefont {M.~A.}\ \bibnamefont
  {Webb}},\ }\href {\doibase 10.1021/acsmacrolett.4c00178} {\bibfield
  {journal} {\bibinfo  {journal} {ACS Macro Lett.}\ }\textbf {\bibinfo {volume}
  {13}},\ \bibinfo {pages} {818} (\bibinfo {year} {2024})}\BibitemShut
  {NoStop}%
\bibitem [{\citenamefont {Chakravarti}\ and\ \citenamefont
  {Joseph}(2025)}]{chakravarti_accurate_2025}%
  \BibitemOpen
  \bibfield  {author} {\bibinfo {author} {\bibfnamefont {A.}~\bibnamefont
  {Chakravarti}}\ and\ \bibinfo {author} {\bibfnamefont {J.~A.}\ \bibnamefont
  {Joseph}},\ }\href {\doibase 10.1002/pro.70284} {\bibfield  {journal}
  {\bibinfo  {journal} {Protein Sci.}\ }\textbf {\bibinfo {volume} {34}},\
  \bibinfo {pages} {e70284} (\bibinfo {year} {2025})}\BibitemShut {NoStop}%
\bibitem [{\citenamefont {Mukherjee}\ and\ \citenamefont
  {Sch{\"a}fer}(2023)}]{mukherjee_therodynamic_2023}%
  \BibitemOpen
  \bibfield  {author} {\bibinfo {author} {\bibfnamefont {S.}~\bibnamefont
  {Mukherjee}}\ and\ \bibinfo {author} {\bibfnamefont {L.~V.}\ \bibnamefont
  {Sch{\"a}fer}},\ }\href {\doibase 10.1038/s41467-023-41586-y} {\bibfield
  {journal} {\bibinfo  {journal} {Nat. Commun.}\ }\textbf {\bibinfo {volume}
  {14}},\ \bibinfo {pages} {5892} (\bibinfo {year} {2023})}\BibitemShut
  {NoStop}%
\bibitem [{\citenamefont {Mukherjee}\ \emph {et~al.}(2024)\citenamefont
  {Mukherjee}, \citenamefont {Ramos}, \citenamefont {Pezzotti}, \citenamefont
  {Kalarikkal}, \citenamefont {Prass}, \citenamefont {Galazzo}, \citenamefont
  {Gendreizig}, \citenamefont {Barbosa}, \citenamefont {Bordignon},
  \citenamefont {Havenith},\ and\ \citenamefont
  {Sch{\"a}fer}}]{mukherjee_entropy_2024}%
  \BibitemOpen
  \bibfield  {author} {\bibinfo {author} {\bibfnamefont {S.}~\bibnamefont
  {Mukherjee}}, \bibinfo {author} {\bibfnamefont {S.}~\bibnamefont {Ramos}},
  \bibinfo {author} {\bibfnamefont {S.}~\bibnamefont {Pezzotti}}, \bibinfo
  {author} {\bibfnamefont {A.}~\bibnamefont {Kalarikkal}}, \bibinfo {author}
  {\bibfnamefont {T.~M.}\ \bibnamefont {Prass}}, \bibinfo {author}
  {\bibfnamefont {L.}~\bibnamefont {Galazzo}}, \bibinfo {author} {\bibfnamefont
  {D.}~\bibnamefont {Gendreizig}}, \bibinfo {author} {\bibfnamefont
  {N.}~\bibnamefont {Barbosa}}, \bibinfo {author} {\bibfnamefont
  {E.}~\bibnamefont {Bordignon}}, \bibinfo {author} {\bibfnamefont
  {M.}~\bibnamefont {Havenith}}, \ and\ \bibinfo {author} {\bibfnamefont
  {L.~V.}\ \bibnamefont {Sch{\"a}fer}},\ }\href {\doibase
  10.1021/acs.jpclett.3c03421} {\bibfield  {journal} {\bibinfo  {journal} {J.
  Phys. Chem. Lett.}\ }\textbf {\bibinfo {volume} {15}},\ \bibinfo {pages}
  {4047} (\bibinfo {year} {2024})}\BibitemShut {NoStop}%
\bibitem [{\citenamefont {Thirumalai}, \citenamefont {Reddy},\ and\
  \citenamefont {Straub}(2012)}]{thirumalai_role_2012}%
  \BibitemOpen
  \bibfield  {author} {\bibinfo {author} {\bibfnamefont {D.}~\bibnamefont
  {Thirumalai}}, \bibinfo {author} {\bibfnamefont {G.}~\bibnamefont {Reddy}}, \
  and\ \bibinfo {author} {\bibfnamefont {J.~E.}\ \bibnamefont {Straub}},\
  }\href {\doibase 10.1021/ar2000869} {\bibfield  {journal} {\bibinfo
  {journal} {Acc. Chem. Res.}\ }\textbf {\bibinfo {volume} {45}},\ \bibinfo
  {pages} {83} (\bibinfo {year} {2012})}\BibitemShut {NoStop}%
\bibitem [{\citenamefont {Ribeiro}\ \emph {et~al.}(2019)\citenamefont
  {Ribeiro}, \citenamefont {Samanta}, \citenamefont {Ebbinghaus},\ and\
  \citenamefont {Marcos}}]{ribeiro_synergic_2019}%
  \BibitemOpen
  \bibfield  {author} {\bibinfo {author} {\bibfnamefont {S.~S.}\ \bibnamefont
  {Ribeiro}}, \bibinfo {author} {\bibfnamefont {N.}~\bibnamefont {Samanta}},
  \bibinfo {author} {\bibfnamefont {S.}~\bibnamefont {Ebbinghaus}}, \ and\
  \bibinfo {author} {\bibfnamefont {J.~C.}\ \bibnamefont {Marcos}},\ }\href
  {\doibase 10.1038/s41570-019-0120-4} {\bibfield  {journal} {\bibinfo
  {journal} {Nat. Rev. Chem.}\ }\textbf {\bibinfo {volume} {3}},\ \bibinfo
  {pages} {552} (\bibinfo {year} {2019})}\BibitemShut {NoStop}%
\bibitem [{\citenamefont {Ahlers}\ \emph {et~al.}(2021)\citenamefont {Ahlers},
  \citenamefont {Adams}, \citenamefont {Bader}, \citenamefont {Pezzotti},
  \citenamefont {Winklhofer}, \citenamefont {Tatzelt},\ and\ \citenamefont
  {Havenith}}]{ahlers_2021}%
  \BibitemOpen
  \bibfield  {author} {\bibinfo {author} {\bibfnamefont {J.}~\bibnamefont
  {Ahlers}}, \bibinfo {author} {\bibfnamefont {E.~M.}\ \bibnamefont {Adams}},
  \bibinfo {author} {\bibfnamefont {V.}~\bibnamefont {Bader}}, \bibinfo
  {author} {\bibfnamefont {S.}~\bibnamefont {Pezzotti}}, \bibinfo {author}
  {\bibfnamefont {K.~F.}\ \bibnamefont {Winklhofer}}, \bibinfo {author}
  {\bibfnamefont {J.}~\bibnamefont {Tatzelt}}, \ and\ \bibinfo {author}
  {\bibfnamefont {M.}~\bibnamefont {Havenith}},\ }\href {\doibase
  10.1016/j.bpj.2021.01.019} {\bibfield  {journal} {\bibinfo  {journal}
  {Biophys. J.}\ }\textbf {\bibinfo {volume} {120}},\ \bibinfo {pages} {1266}
  (\bibinfo {year} {2021})}\BibitemShut {NoStop}%
\bibitem [{\citenamefont {Pezzotti}\ \emph {et~al.}(2023)\citenamefont
  {Pezzotti}, \citenamefont {K{\"o}nig}, \citenamefont {Ramos}, \citenamefont
  {Schwaab},\ and\ \citenamefont {Havenith}}]{pezzotti_2023}%
  \BibitemOpen
  \bibfield  {author} {\bibinfo {author} {\bibfnamefont {S.}~\bibnamefont
  {Pezzotti}}, \bibinfo {author} {\bibfnamefont {B.}~\bibnamefont {K{\"o}nig}},
  \bibinfo {author} {\bibfnamefont {S.}~\bibnamefont {Ramos}}, \bibinfo
  {author} {\bibfnamefont {G.}~\bibnamefont {Schwaab}}, \ and\ \bibinfo
  {author} {\bibfnamefont {M.}~\bibnamefont {Havenith}},\ }\href {\doibase
  10.1021/acs.jpclett.2c02697} {\bibfield  {journal} {\bibinfo  {journal} {J.
  Phys. Chem. Lett.}\ }\textbf {\bibinfo {volume} {14}},\ \bibinfo {pages}
  {1556} (\bibinfo {year} {2023})}\BibitemShut {NoStop}%
\bibitem [{\citenamefont {Garaizar}\ \emph {et~al.}(2020)\citenamefont
  {Garaizar}, \citenamefont {Sanchez-Burgos}, \citenamefont
  {Collepardo-Guevara},\ and\ \citenamefont
  {Espinosa}}]{garaizar_expansion_2020}%
  \BibitemOpen
  \bibfield  {author} {\bibinfo {author} {\bibfnamefont {A.}~\bibnamefont
  {Garaizar}}, \bibinfo {author} {\bibfnamefont {I.}~\bibnamefont
  {Sanchez-Burgos}}, \bibinfo {author} {\bibfnamefont {R.}~\bibnamefont
  {Collepardo-Guevara}}, \ and\ \bibinfo {author} {\bibfnamefont {J.~R.}\
  \bibnamefont {Espinosa}},\ }\href {\doibase 10.3390/molecules25204705}
  {\bibfield  {journal} {\bibinfo  {journal} {Molecules}\ }\textbf {\bibinfo
  {volume} {25}},\ \bibinfo {pages} {4705} (\bibinfo {year}
  {2020})}\BibitemShut {NoStop}%
\bibitem [{\citenamefont {Rovigatti}\ and\ \citenamefont
  {Sciortino}(2022)}]{rovigatti_designing_2022}%
  \BibitemOpen
  \bibfield  {author} {\bibinfo {author} {\bibfnamefont {L.}~\bibnamefont
  {Rovigatti}}\ and\ \bibinfo {author} {\bibfnamefont {F.}~\bibnamefont
  {Sciortino}},\ }\href {\doibase 10.1103/physrevlett.129.047801} {\bibfield
  {journal} {\bibinfo  {journal} {Phys. Rev. Lett.}\ }\textbf {\bibinfo
  {volume} {112}},\ \bibinfo {pages} {047801} (\bibinfo {year}
  {2022})}\BibitemShut {NoStop}%
\bibitem [{\citenamefont {Rovigatti}\ and\ \citenamefont
  {Sciortino}(2023)}]{rovigatti_entropy-driven_2023}%
  \BibitemOpen
  \bibfield  {author} {\bibinfo {author} {\bibfnamefont {L.}~\bibnamefont
  {Rovigatti}}\ and\ \bibinfo {author} {\bibfnamefont {F.}~\bibnamefont
  {Sciortino}},\ }\href {\doibase 10.21468/SciPostPhys.15.4.163} {\bibfield
  {journal} {\bibinfo  {journal} {SciPost Physics}\ }\textbf {\bibinfo {volume}
  {15}},\ \bibinfo {pages} {163} (\bibinfo {year} {2023})}\BibitemShut
  {NoStop}%
\bibitem [{\citenamefont {Boeynaems}\ \emph {et~al.}(2018)\citenamefont
  {Boeynaems}, \citenamefont {Alberti}, \citenamefont {Fawzi}, \citenamefont
  {Mittag}, \citenamefont {Polymenidou}, \citenamefont {Rousseau},
  \citenamefont {Schymkowitz}, \citenamefont {Shorter}, \citenamefont
  {Wolozin}, \citenamefont {Van Den~Bosch}, \citenamefont {Tompa},\ and\
  \citenamefont {Fuxreiter}}]{boeynaems_protein_2018}%
  \BibitemOpen
  \bibfield  {author} {\bibinfo {author} {\bibfnamefont {S.}~\bibnamefont
  {Boeynaems}}, \bibinfo {author} {\bibfnamefont {S.}~\bibnamefont {Alberti}},
  \bibinfo {author} {\bibfnamefont {N.~L.}\ \bibnamefont {Fawzi}}, \bibinfo
  {author} {\bibfnamefont {T.}~\bibnamefont {Mittag}}, \bibinfo {author}
  {\bibfnamefont {M.}~\bibnamefont {Polymenidou}}, \bibinfo {author}
  {\bibfnamefont {F.}~\bibnamefont {Rousseau}}, \bibinfo {author}
  {\bibfnamefont {J.}~\bibnamefont {Schymkowitz}}, \bibinfo {author}
  {\bibfnamefont {J.}~\bibnamefont {Shorter}}, \bibinfo {author} {\bibfnamefont
  {B.}~\bibnamefont {Wolozin}}, \bibinfo {author} {\bibfnamefont
  {L.}~\bibnamefont {Van Den~Bosch}}, \bibinfo {author} {\bibfnamefont
  {P.}~\bibnamefont {Tompa}}, \ and\ \bibinfo {author} {\bibfnamefont
  {M.}~\bibnamefont {Fuxreiter}},\ }\href {\doibase 10.1016/j.tcb.2018.02.004}
  {\bibfield  {journal} {\bibinfo  {journal} {Trends Cell Biol.}\ }\textbf
  {\bibinfo {volume} {28}},\ \bibinfo {pages} {420} (\bibinfo {year}
  {2018})}\BibitemShut {NoStop}%
\bibitem [{\citenamefont {Kim}\ and\ \citenamefont
  {Matsunaga}(2017)}]{kim_thermo-responsive_2017}%
  \BibitemOpen
  \bibfield  {author} {\bibinfo {author} {\bibfnamefont {Y.-J.}\ \bibnamefont
  {Kim}}\ and\ \bibinfo {author} {\bibfnamefont {Y.~T.}\ \bibnamefont
  {Matsunaga}},\ }\href {\doibase 10.1039/C7TB00157F} {\bibfield  {journal}
  {\bibinfo  {journal} {J. Mater. Chem. B.}\ }\textbf {\bibinfo {volume} {5}},\
  \bibinfo {pages} {4307} (\bibinfo {year} {2017})}\BibitemShut {NoStop}%
\bibitem [{\citenamefont {Dai}, \citenamefont {You},\ and\ \citenamefont
  {Chilkoti}(2023)}]{dai_engineering_2023}%
  \BibitemOpen
  \bibfield  {author} {\bibinfo {author} {\bibfnamefont {Y.}~\bibnamefont
  {Dai}}, \bibinfo {author} {\bibfnamefont {L.}~\bibnamefont {You}}, \ and\
  \bibinfo {author} {\bibfnamefont {A.}~\bibnamefont {Chilkoti}},\ }\href
  {\doibase 10.1038/s44222-023-00052-6} {\bibfield  {journal} {\bibinfo
  {journal} {Nat. Rev. Bioeng.}\ }\textbf {\bibinfo {volume} {1}},\ \bibinfo
  {pages} {466} (\bibinfo {year} {2023})}\BibitemShut {NoStop}%
\bibitem [{\citenamefont {Tamaki}\ and\ \citenamefont
  {Kojima}(2020)}]{tamaki_ph-switchable_2020}%
  \BibitemOpen
  \bibfield  {author} {\bibinfo {author} {\bibfnamefont {M.}~\bibnamefont
  {Tamaki}}\ and\ \bibinfo {author} {\bibfnamefont {C.}~\bibnamefont
  {Kojima}},\ }\href {\doibase 10.1039/D0RA00499E} {\bibfield  {journal}
  {\bibinfo  {journal} {RSC Adv.}\ }\textbf {\bibinfo {volume} {10}},\ \bibinfo
  {pages} {10452} (\bibinfo {year} {2020})}\BibitemShut {NoStop}%
\bibitem [{\citenamefont {Tangade}\ \emph {et~al.}(2025)\citenamefont
  {Tangade}, \citenamefont {Mondal}, \citenamefont {Wang}, \citenamefont
  {Kim},\ and\ \citenamefont {Mittal}}]{tangade_multiphasic_2025}%
  \BibitemOpen
  \bibfield  {author} {\bibinfo {author} {\bibfnamefont {A.~S.}\ \bibnamefont
  {Tangade}}, \bibinfo {author} {\bibfnamefont {A.}~\bibnamefont {Mondal}},
  \bibinfo {author} {\bibfnamefont {J.}~\bibnamefont {Wang}}, \bibinfo {author}
  {\bibfnamefont {Y.~C.}\ \bibnamefont {Kim}}, \ and\ \bibinfo {author}
  {\bibfnamefont {J.}~\bibnamefont {Mittal}},\ }\href {\doibase
  10.1021/acs.jpcb.5c03987} {\bibfield  {journal} {\bibinfo  {journal} {J.
  Phys. Chem. B}\ }\textbf {\bibinfo {volume} {129}},\ \bibinfo {pages} {9588}
  (\bibinfo {year} {2025})}\BibitemShut {NoStop}%
\bibitem [{\citenamefont {Roy}, \citenamefont {Santra},\ and\ \citenamefont
  {Singh}(2025)}]{roy_anomalous_2025}%
  \BibitemOpen
  \bibfield  {author} {\bibinfo {author} {\bibfnamefont {S.}~\bibnamefont
  {Roy}}, \bibinfo {author} {\bibfnamefont {M.}~\bibnamefont {Santra}}, \ and\
  \bibinfo {author} {\bibfnamefont {R.~S.}\ \bibnamefont {Singh}},\ }\href
  {\doibase 10.1063/5.0275549} {\bibfield  {journal} {\bibinfo  {journal} {J.
  Chem. Phys.}\ }\textbf {\bibinfo {volume} {163}},\ \bibinfo {pages} {024905}
  (\bibinfo {year} {2025})}\BibitemShut {NoStop}%
\bibitem [{\citenamefont {Lichtinger}\ \emph {et~al.}(2021)\citenamefont
  {Lichtinger}, \citenamefont {Garaizar}, \citenamefont {Collepardo-Guevara},\
  and\ \citenamefont {Reinhardt}}]{lichtinger_targeted_2021}%
  \BibitemOpen
  \bibfield  {author} {\bibinfo {author} {\bibfnamefont {S.~M.}\ \bibnamefont
  {Lichtinger}}, \bibinfo {author} {\bibfnamefont {A.}~\bibnamefont
  {Garaizar}}, \bibinfo {author} {\bibfnamefont {R.}~\bibnamefont
  {Collepardo-Guevara}}, \ and\ \bibinfo {author} {\bibfnamefont
  {A.}~\bibnamefont {Reinhardt}},\ }\href {\doibase
  10.1371/journal.pcbi.1009328} {\bibfield  {journal} {\bibinfo  {journal}
  {PLOS Comput. Biol.}\ }\textbf {\bibinfo {volume} {17}},\ \bibinfo {pages}
  {e1009328} (\bibinfo {year} {2021})}\BibitemShut {NoStop}%
\bibitem [{\citenamefont {Sahli}\ \emph {et~al.}(2019)\citenamefont {Sahli},
  \citenamefont {Renard}, \citenamefont {Solé-Jamault}, \citenamefont
  {Giuliani},\ and\ \citenamefont {Boire}}]{sahli_role_2019}%
  \BibitemOpen
  \bibfield  {author} {\bibinfo {author} {\bibfnamefont {L.}~\bibnamefont
  {Sahli}}, \bibinfo {author} {\bibfnamefont {D.}~\bibnamefont {Renard}},
  \bibinfo {author} {\bibfnamefont {V.}~\bibnamefont {Solé-Jamault}}, \bibinfo
  {author} {\bibfnamefont {A.}~\bibnamefont {Giuliani}}, \ and\ \bibinfo
  {author} {\bibfnamefont {A.}~\bibnamefont {Boire}},\ }\href {\doibase
  10.1038/s41598-019-49745-2} {\bibfield  {journal} {\bibinfo  {journal} {Sci.
  Rep.}\ }\textbf {\bibinfo {volume} {9}},\ \bibinfo {pages} {13391} (\bibinfo
  {year} {2019})}\BibitemShut {NoStop}%
\bibitem [{\citenamefont {Nikfarjam}\ \emph {et~al.}(2020)\citenamefont
  {Nikfarjam}, \citenamefont {Jouravleva}, \citenamefont {Anisimov},\ and\
  \citenamefont {Woehl}}]{nikfarjam_effects_2020}%
  \BibitemOpen
  \bibfield  {author} {\bibinfo {author} {\bibfnamefont {S.}~\bibnamefont
  {Nikfarjam}}, \bibinfo {author} {\bibfnamefont {E.~V.}\ \bibnamefont
  {Jouravleva}}, \bibinfo {author} {\bibfnamefont {M.~A.}\ \bibnamefont
  {Anisimov}}, \ and\ \bibinfo {author} {\bibfnamefont {T.~J.}\ \bibnamefont
  {Woehl}},\ }\href {\doibase 10.3390/biom10091262} {\bibfield  {journal}
  {\bibinfo  {journal} {Biomolecules}\ }\textbf {\bibinfo {volume} {10}},\
  \bibinfo {pages} {1262} (\bibinfo {year} {2020})}\BibitemShut {NoStop}%
\bibitem [{\citenamefont {Zeng}\ \emph {et~al.}(2020)\citenamefont {Zeng},
  \citenamefont {Holehouse}, \citenamefont {Chilkoti}, \citenamefont {Mittag},\
  and\ \citenamefont {Pappu}}]{zeng_connecting_2020}%
  \BibitemOpen
  \bibfield  {author} {\bibinfo {author} {\bibfnamefont {X.}~\bibnamefont
  {Zeng}}, \bibinfo {author} {\bibfnamefont {A.~S.}\ \bibnamefont {Holehouse}},
  \bibinfo {author} {\bibfnamefont {A.}~\bibnamefont {Chilkoti}}, \bibinfo
  {author} {\bibfnamefont {T.}~\bibnamefont {Mittag}}, \ and\ \bibinfo {author}
  {\bibfnamefont {R.~V.}\ \bibnamefont {Pappu}},\ }\href {\doibase
  10.1016/j.bpj.2020.06.014} {\bibfield  {journal} {\bibinfo  {journal}
  {Biophys. J.}\ }\textbf {\bibinfo {volume} {119}},\ \bibinfo {pages} {402}
  (\bibinfo {year} {2020})}\BibitemShut {NoStop}%
\bibitem [{\citenamefont {Dannenhoffer-Lafage}\ and\ \citenamefont
  {Best}(2021)}]{dannenhoffer-lafage_data-driven_2021}%
  \BibitemOpen
  \bibfield  {author} {\bibinfo {author} {\bibfnamefont {T.}~\bibnamefont
  {Dannenhoffer-Lafage}}\ and\ \bibinfo {author} {\bibfnamefont {R.~B.}\
  \bibnamefont {Best}},\ }\href {\doibase 10.1021/acs.jpcb.0c11479} {\bibfield
  {journal} {\bibinfo  {journal} {J. Phys. Chem. B}\ }\textbf {\bibinfo
  {volume} {125}},\ \bibinfo {pages} {4046} (\bibinfo {year}
  {2021})}\BibitemShut {NoStop}%
\bibitem [{\citenamefont {Majumdar}\ \emph {et~al.}(2019)\citenamefont
  {Majumdar}, \citenamefont {Dogra}, \citenamefont {Maity},\ and\ \citenamefont
  {Mukhopadhyay}}]{doi:10.1021/acs.jpclett.9b01731}%
  \BibitemOpen
  \bibfield  {author} {\bibinfo {author} {\bibfnamefont {A.}~\bibnamefont
  {Majumdar}}, \bibinfo {author} {\bibfnamefont {P.}~\bibnamefont {Dogra}},
  \bibinfo {author} {\bibfnamefont {S.}~\bibnamefont {Maity}}, \ and\ \bibinfo
  {author} {\bibfnamefont {S.}~\bibnamefont {Mukhopadhyay}},\ }\href {\doibase
  10.1021/acs.jpclett.9b01731} {\bibfield  {journal} {\bibinfo  {journal} {J.
  Phys. Chem. Lett.}\ }\textbf {\bibinfo {volume} {10}},\ \bibinfo {pages}
  {3929} (\bibinfo {year} {2019})}\BibitemShut {NoStop}%
\bibitem [{\citenamefont {Hazra}\ and\ \citenamefont
  {Levy}(2021)}]{hazra_biophysics_2021}%
  \BibitemOpen
  \bibfield  {author} {\bibinfo {author} {\bibfnamefont {M.~K.}\ \bibnamefont
  {Hazra}}\ and\ \bibinfo {author} {\bibfnamefont {Y.}~\bibnamefont {Levy}},\
  }\href {\doibase 10.1021/acs.jpcb.0c09975} {\bibfield  {journal} {\bibinfo
  {journal} {J. Phys. Chem. B}\ }\textbf {\bibinfo {volume} {125}},\ \bibinfo
  {pages} {2202} (\bibinfo {year} {2021})}\BibitemShut {NoStop}%
\bibitem [{\citenamefont {Kaur}\ \emph {et~al.}(2021)\citenamefont {Kaur},
  \citenamefont {Raju}, \citenamefont {Alshareedah}, \citenamefont {Davis},
  \citenamefont {Potoyan},\ and\ \citenamefont
  {Banerjee}}]{kaur_sequence-encoded_2021}%
  \BibitemOpen
  \bibfield  {author} {\bibinfo {author} {\bibfnamefont {T.}~\bibnamefont
  {Kaur}}, \bibinfo {author} {\bibfnamefont {M.}~\bibnamefont {Raju}}, \bibinfo
  {author} {\bibfnamefont {I.}~\bibnamefont {Alshareedah}}, \bibinfo {author}
  {\bibfnamefont {R.~B.}\ \bibnamefont {Davis}}, \bibinfo {author}
  {\bibfnamefont {D.~A.}\ \bibnamefont {Potoyan}}, \ and\ \bibinfo {author}
  {\bibfnamefont {P.~R.}\ \bibnamefont {Banerjee}},\ }\href {\doibase
  10.1038/s41467-021-21089-4} {\bibfield  {journal} {\bibinfo  {journal} {Nat.
  Commun.}\ }\textbf {\bibinfo {volume} {12}},\ \bibinfo {pages} {872}
  (\bibinfo {year} {2021})}\BibitemShut {NoStop}%
\bibitem [{\citenamefont {Dignon}\ \emph
  {et~al.}(2018{\natexlab{b}})\citenamefont {Dignon}, \citenamefont {Zheng},
  \citenamefont {Best}, \citenamefont {Kim},\ and\ \citenamefont
  {Mittal}}]{dignon_relation_2018}%
  \BibitemOpen
  \bibfield  {author} {\bibinfo {author} {\bibfnamefont {G.~L.}\ \bibnamefont
  {Dignon}}, \bibinfo {author} {\bibfnamefont {W.}~\bibnamefont {Zheng}},
  \bibinfo {author} {\bibfnamefont {R.~B.}\ \bibnamefont {Best}}, \bibinfo
  {author} {\bibfnamefont {Y.~C.}\ \bibnamefont {Kim}}, \ and\ \bibinfo
  {author} {\bibfnamefont {J.}~\bibnamefont {Mittal}},\ }\href {\doibase
  10.1073/pnas.1804177115} {\bibfield  {journal} {\bibinfo  {journal} {Proc.
  Natl. Acad. Sci. U.S.A.}\ }\textbf {\bibinfo {volume} {115}},\ \bibinfo
  {pages} {9929} (\bibinfo {year} {2018}{\natexlab{b}})}\BibitemShut {NoStop}%
\bibitem [{\citenamefont {Caupin}\ and\ \citenamefont
  {Anisimov}(2021)}]{caupin_minimal_2021}%
  \BibitemOpen
  \bibfield  {author} {\bibinfo {author} {\bibfnamefont {F.}~\bibnamefont
  {Caupin}}\ and\ \bibinfo {author} {\bibfnamefont {M.~A.}\ \bibnamefont
  {Anisimov}},\ }\href {\doibase 10.1103/PhysRevLett.127.185701} {\bibfield
  {journal} {\bibinfo  {journal} {Phys. Rev. Lett.}\ }\textbf {\bibinfo
  {volume} {127}},\ \bibinfo {pages} {185701} (\bibinfo {year}
  {2021})}\BibitemShut {NoStop}%
\bibitem [{\citenamefont {Longo}\ \emph {et~al.}(2022)\citenamefont {Longo},
  \citenamefont {Shumovskyi}, \citenamefont {Asadov}, \citenamefont
  {Buldryev},\ and\ \citenamefont {Anisimov}}]{longo_structure_2022}%
  \BibitemOpen
  \bibfield  {author} {\bibinfo {author} {\bibfnamefont {T.~J.}\ \bibnamefont
  {Longo}}, \bibinfo {author} {\bibfnamefont {N.~A.}\ \bibnamefont
  {Shumovskyi}}, \bibinfo {author} {\bibfnamefont {S.~M.}\ \bibnamefont
  {Asadov}}, \bibinfo {author} {\bibfnamefont {S.~V.}\ \bibnamefont
  {Buldryev}}, \ and\ \bibinfo {author} {\bibfnamefont {M.~A.}\ \bibnamefont
  {Anisimov}},\ }\href {\doibase 10.1016/j.nocx.2022.100082} {\bibfield
  {journal} {\bibinfo  {journal} {J. Non-Cryst. Solids: X}\ }\textbf {\bibinfo
  {volume} {13}},\ \bibinfo {pages} {100082} (\bibinfo {year}
  {2022})}\BibitemShut {NoStop}%
\bibitem [{\citenamefont {Longo}\ and\ \citenamefont
  {Anisimov}(2022)}]{longo_phase_2022}%
  \BibitemOpen
  \bibfield  {author} {\bibinfo {author} {\bibfnamefont {T.~J.}\ \bibnamefont
  {Longo}}\ and\ \bibinfo {author} {\bibfnamefont {M.~A.}\ \bibnamefont
  {Anisimov}},\ }\href {\doibase 10.1063/5.0081180} {\bibfield  {journal}
  {\bibinfo  {journal} {J. Chem. Phys.}\ }\textbf {\bibinfo {volume} {156}},\
  \bibinfo {pages} {084502} (\bibinfo {year} {2022})}\BibitemShut {NoStop}%
\bibitem [{\citenamefont {Longo}\ \emph {et~al.}(2023)\citenamefont {Longo},
  \citenamefont {Shumovskyi}, \citenamefont {Uralcan}, \citenamefont
  {Buldyrev}, \citenamefont {Anisimov},\ and\ \citenamefont
  {Debenedetti}}]{longo_formation_2023}%
  \BibitemOpen
  \bibfield  {author} {\bibinfo {author} {\bibfnamefont {T.~J.}\ \bibnamefont
  {Longo}}, \bibinfo {author} {\bibfnamefont {N.~A.}\ \bibnamefont
  {Shumovskyi}}, \bibinfo {author} {\bibfnamefont {B.}~\bibnamefont {Uralcan}},
  \bibinfo {author} {\bibfnamefont {S.~V.}\ \bibnamefont {Buldyrev}}, \bibinfo
  {author} {\bibfnamefont {M.~A.}\ \bibnamefont {Anisimov}}, \ and\ \bibinfo
  {author} {\bibfnamefont {P.~G.}\ \bibnamefont {Debenedetti}},\ }\href
  {\doibase 10.1073/pnas.2215012120} {\bibfield  {journal} {\bibinfo  {journal}
  {Proc. Natl. Acad. Sci. U.S.A.}\ }\textbf {\bibinfo {volume} {120}},\
  \bibinfo {pages} {e2215012120} (\bibinfo {year} {2023})}\BibitemShut
  {NoStop}%
\bibitem [{\citenamefont {Cho}\ and\ \citenamefont
  {Jacobs}(2023{\natexlab{a}})}]{cho_tuning_2023}%
  \BibitemOpen
  \bibfield  {author} {\bibinfo {author} {\bibfnamefont {Y.}~\bibnamefont
  {Cho}}\ and\ \bibinfo {author} {\bibfnamefont {W.~M.}\ \bibnamefont
  {Jacobs}},\ }\href {\doibase 10.1103/PhysRevLett.130.128203} {\bibfield
  {journal} {\bibinfo  {journal} {Phys. Rev. Lett.}\ }\textbf {\bibinfo
  {volume} {130}},\ \bibinfo {pages} {128203} (\bibinfo {year}
  {2023}{\natexlab{a}})}\BibitemShut {NoStop}%
\bibitem [{\citenamefont {Cho}\ and\ \citenamefont
  {Jacobs}(2023{\natexlab{b}})}]{cho_nonequilibrium_2023}%
  \BibitemOpen
  \bibfield  {author} {\bibinfo {author} {\bibfnamefont {Y.}~\bibnamefont
  {Cho}}\ and\ \bibinfo {author} {\bibfnamefont {W.~M.}\ \bibnamefont
  {Jacobs}},\ }\href {\doibase 10.1063/5.0166824} {\bibfield  {journal}
  {\bibinfo  {journal} {J. Chem. Phys.}\ }\textbf {\bibinfo {volume} {159}},\
  \bibinfo {pages} {154101} (\bibinfo {year} {2023}{\natexlab{b}})}\BibitemShut
  {NoStop}%
\bibitem [{\citenamefont {Flory}(1942)}]{flory_thermodynamics_1942}%
  \BibitemOpen
  \bibfield  {author} {\bibinfo {author} {\bibfnamefont {P.~J.}\ \bibnamefont
  {Flory}},\ }\href {\doibase 10.1063/1.1723621} {\bibfield  {journal}
  {\bibinfo  {journal} {J. Chem. Phys.}\ }\textbf {\bibinfo {volume} {10}},\
  \bibinfo {pages} {51} (\bibinfo {year} {1942})}\BibitemShut {NoStop}%
\bibitem [{\citenamefont {Huggins}(1942)}]{huggins_thermodynamic_1942}%
  \BibitemOpen
  \bibfield  {author} {\bibinfo {author} {\bibfnamefont {M.~L.}\ \bibnamefont
  {Huggins}},\ }\href {\doibase 10.1111/j.1749-6632.1942.tb47940.x} {\bibfield
  {journal} {\bibinfo  {journal} {Ann. N.Y. Acad. Sci.}\ }\textbf {\bibinfo
  {volume} {43}},\ \bibinfo {pages} {1} (\bibinfo {year} {1942})}\BibitemShut
  {NoStop}%
\bibitem [{\citenamefont {Kawasaki}(1966)}]{PhysRev.145.224}%
  \BibitemOpen
  \bibfield  {author} {\bibinfo {author} {\bibfnamefont {K.}~\bibnamefont
  {Kawasaki}},\ }\href {\doibase 10.1103/PhysRev.145.224} {\bibfield  {journal}
  {\bibinfo  {journal} {Phys. Rev.}\ }\textbf {\bibinfo {volume} {145}},\
  \bibinfo {pages} {224} (\bibinfo {year} {1966})}\BibitemShut {NoStop}%
\bibitem [{\citenamefont {Frenkel}\ and\ \citenamefont
  {Smit}(2023)}]{frenkel_understanding_2023}%
  \BibitemOpen
  \bibfield  {author} {\bibinfo {author} {\bibfnamefont {D.}~\bibnamefont
  {Frenkel}}\ and\ \bibinfo {author} {\bibfnamefont {B.}~\bibnamefont {Smit}},\
  }\href@noop {} {\emph {\bibinfo {title} {Understanding Molecular Simulation:
  From Algorithms to Applications}}},\ \bibinfo {edition} {3rd}\ ed.\ (\bibinfo
   {publisher} {Elsevier},\ \bibinfo {year} {2023})\BibitemShut {NoStop}%
\bibitem [{\citenamefont {Chauhan}\ \emph {et~al.}(2025)\citenamefont
  {Chauhan}, \citenamefont {Wilkinson}, \citenamefont {Yuan}, \citenamefont
  {Cohen}, \citenamefont {Onishi}, \citenamefont {Pappu},\ and\ \citenamefont
  {Strader}}]{chauhan_active_2025}%
  \BibitemOpen
  \bibfield  {author} {\bibinfo {author} {\bibfnamefont {G.}~\bibnamefont
  {Chauhan}}, \bibinfo {author} {\bibfnamefont {E.~G.}\ \bibnamefont
  {Wilkinson}}, \bibinfo {author} {\bibfnamefont {Y.}~\bibnamefont {Yuan}},
  \bibinfo {author} {\bibfnamefont {S.~R.}\ \bibnamefont {Cohen}}, \bibinfo
  {author} {\bibfnamefont {M.}~\bibnamefont {Onishi}}, \bibinfo {author}
  {\bibfnamefont {R.~V.}\ \bibnamefont {Pappu}}, \ and\ \bibinfo {author}
  {\bibfnamefont {L.~C.}\ \bibnamefont {Strader}},\ }\href {\doibase
  10.1126/sciadv.adv7875} {\bibfield  {journal} {\bibinfo  {journal} {Sci.
  Adv.}\ }\textbf {\bibinfo {volume} {11}},\ \bibinfo {pages} {eadv7875}
  (\bibinfo {year} {2025})}\BibitemShut {NoStop}%
\bibitem [{\citenamefont {Mizuno}\ \emph {et~al.}(2007)\citenamefont {Mizuno},
  \citenamefont {Tardin}, \citenamefont {Schmidt},\ and\ \citenamefont
  {MacKintosh}}]{mizuno_nonequilibrium_2007}%
  \BibitemOpen
  \bibfield  {author} {\bibinfo {author} {\bibfnamefont {D.}~\bibnamefont
  {Mizuno}}, \bibinfo {author} {\bibfnamefont {C.}~\bibnamefont {Tardin}},
  \bibinfo {author} {\bibfnamefont {C.~F.}\ \bibnamefont {Schmidt}}, \ and\
  \bibinfo {author} {\bibfnamefont {F.~C.}\ \bibnamefont {MacKintosh}},\ }\href
  {\doibase 10.1126/science.1134404} {\bibfield  {journal} {\bibinfo  {journal}
  {Science}\ }\textbf {\bibinfo {volume} {315}},\ \bibinfo {pages} {370}
  (\bibinfo {year} {2007})}\BibitemShut {NoStop}%
\bibitem [{\citenamefont {Fakhri}\ \emph {et~al.}(2014)\citenamefont {Fakhri},
  \citenamefont {Wessel}, \citenamefont {Willms}, \citenamefont {Pasquali},
  \citenamefont {Klopfenstein}, \citenamefont {MacKintosh},\ and\ \citenamefont
  {Schmidt}}]{fakhri_high-resolution_2014}%
  \BibitemOpen
  \bibfield  {author} {\bibinfo {author} {\bibfnamefont {N.}~\bibnamefont
  {Fakhri}}, \bibinfo {author} {\bibfnamefont {A.~D.}\ \bibnamefont {Wessel}},
  \bibinfo {author} {\bibfnamefont {C.}~\bibnamefont {Willms}}, \bibinfo
  {author} {\bibfnamefont {M.}~\bibnamefont {Pasquali}}, \bibinfo {author}
  {\bibfnamefont {D.~R.}\ \bibnamefont {Klopfenstein}}, \bibinfo {author}
  {\bibfnamefont {F.~C.}\ \bibnamefont {MacKintosh}}, \ and\ \bibinfo {author}
  {\bibfnamefont {C.~F.}\ \bibnamefont {Schmidt}},\ }\href {\doibase
  10.1126/science.1250170} {\bibfield  {journal} {\bibinfo  {journal}
  {Science}\ }\textbf {\bibinfo {volume} {344}},\ \bibinfo {pages} {1031}
  (\bibinfo {year} {2014})}\BibitemShut {NoStop}%
\bibitem [{\citenamefont {Wang}\ \emph {et~al.}(2025)\citenamefont {Wang},
  \citenamefont {Nikoubashman}, \citenamefont {Kim},\ and\ \citenamefont
  {Mittal}}]{wang_controlling_2025}%
  \BibitemOpen
  \bibfield  {author} {\bibinfo {author} {\bibfnamefont {J.}~\bibnamefont
  {Wang}}, \bibinfo {author} {\bibfnamefont {A.}~\bibnamefont {Nikoubashman}},
  \bibinfo {author} {\bibfnamefont {Y.~C.}\ \bibnamefont {Kim}}, \ and\
  \bibinfo {author} {\bibfnamefont {J.}~\bibnamefont {Mittal}},\ }\href
  {\doibase 10.1021/jacsau.5c00713} {\bibfield  {journal} {\bibinfo  {journal}
  {JACS Au}\ }\textbf {\bibinfo {volume} {5}},\ \bibinfo {pages} {4064}
  (\bibinfo {year} {2025})}\BibitemShut {NoStop}%
\bibitem [{\citenamefont {Rana}\ \emph {et~al.}(2024)\citenamefont {Rana},
  \citenamefont {Xu}, \citenamefont {Narayanan}, \citenamefont {Walls},
  \citenamefont {Panagiotopoulos}, \citenamefont {Avalos},\ and\ \citenamefont
  {Brangwynne}}]{rana_asymmetric_2024}%
  \BibitemOpen
  \bibfield  {author} {\bibinfo {author} {\bibfnamefont {U.}~\bibnamefont
  {Rana}}, \bibinfo {author} {\bibfnamefont {K.}~\bibnamefont {Xu}}, \bibinfo
  {author} {\bibfnamefont {A.}~\bibnamefont {Narayanan}}, \bibinfo {author}
  {\bibfnamefont {M.~T.}\ \bibnamefont {Walls}}, \bibinfo {author}
  {\bibfnamefont {A.~Z.}\ \bibnamefont {Panagiotopoulos}}, \bibinfo {author}
  {\bibfnamefont {J.~L.}\ \bibnamefont {Avalos}}, \ and\ \bibinfo {author}
  {\bibfnamefont {C.~P.}\ \bibnamefont {Brangwynne}},\ }\href {\doibase
  10.1038/s41557-024-01456-6} {\bibfield  {journal} {\bibinfo  {journal} {Nat.
  Chem.}\ }\textbf {\bibinfo {volume} {16}},\ \bibinfo {pages} {1073} (\bibinfo
  {year} {2024})}\BibitemShut {NoStop}%
\bibitem [{\citenamefont {Ramachandran}\ and\ \citenamefont
  {Potoyan}(2026)}]{Ramachandran2026.03.04.709570}%
  \BibitemOpen
  \bibfield  {author} {\bibinfo {author} {\bibfnamefont {V.}~\bibnamefont
  {Ramachandran}}\ and\ \bibinfo {author} {\bibfnamefont {D.~A.}\ \bibnamefont
  {Potoyan}},\ }\href {\doibase 10.64898/2026.03.04.709570} {\bibfield
  {journal} {\bibinfo  {journal} {bioRxiv}\ } (\bibinfo {year} {2026}),\
  10.64898/2026.03.04.709570}\BibitemShut {NoStop}%
\bibitem [{\citenamefont {Rauh}, \citenamefont {Tesei},\ and\ \citenamefont
  {Lindorff‐Larsen}(2025)}]{rauh_coarsegrained_2025}%
  \BibitemOpen
  \bibfield  {author} {\bibinfo {author} {\bibfnamefont {A.~S.}\ \bibnamefont
  {Rauh}}, \bibinfo {author} {\bibfnamefont {G.}~\bibnamefont {Tesei}}, \ and\
  \bibinfo {author} {\bibfnamefont {K.}~\bibnamefont {Lindorff‐Larsen}},\
  }\href {\doibase 10.1002/pro.70232} {\bibfield  {journal} {\bibinfo
  {journal} {Protein Sci.}\ }\textbf {\bibinfo {volume} {34}},\ \bibinfo
  {pages} {e70232} (\bibinfo {year} {2025})}\BibitemShut {NoStop}%
\bibitem [{\citenamefont {Potts}(1952)}]{potts_generalized_1952}%
  \BibitemOpen
  \bibfield  {author} {\bibinfo {author} {\bibfnamefont {R.~B.}\ \bibnamefont
  {Potts}},\ }\href {\doibase 10.1017/S0305004100027419} {\bibfield  {journal}
  {\bibinfo  {journal} {Math. Proc. Cambridge Philos. Soc.}\ }\textbf {\bibinfo
  {volume} {48}},\ \bibinfo {pages} {106} (\bibinfo {year} {1952})}\BibitemShut
  {NoStop}%
\bibitem [{\citenamefont {Wu}(1982)}]{wu_potts_1982}%
  \BibitemOpen
  \bibfield  {author} {\bibinfo {author} {\bibfnamefont {F.~Y.}\ \bibnamefont
  {Wu}},\ }\href {\doibase 10.1103/RevModPhys.54.235} {\bibfield  {journal}
  {\bibinfo  {journal} {Rev. Mod. Phys.}\ }\textbf {\bibinfo {volume} {54}},\
  \bibinfo {pages} {235} (\bibinfo {year} {1982})}\BibitemShut {NoStop}%
\end{thebibliography}%

\end{document}